\documentclass[journal, draftcls, onecolumn]{IEEEtran}
\usepackage{amssymb,amsmath,graphicx,cite,graphics,mathrsfs}
\usepackage{latexsym,amssymb,amsmath,graphicx,cite,bbm,float,graphics}
\usepackage{amsthm}

\newcommand{\be}{\begin{equation}}
\newcommand{\ee}{\end{equation}}
\newcommand{\bea}{\begin{eqnarray}}
\newcommand{\eea}{\end{eqnarray}}
\newcommand{\nn}{\nonumber}

\newcommand{\ei}{\end{itemize}}
\newcommand{\bi}{\begin{itemize}}

\newcommand{\vY}{\mathbf{Y}^{k}}
\newcommand{\vy}{\mathbf{y}^{k}}

\newcommand{\vUo}{\mathbf{U}_{1}^{k}}
\newcommand{\vUt}{\mathbf{U}_{K}^{k}}

\newtheorem{proposition}{Proposition}
\newtheorem{theorem}{Theorem}

\newtheorem{lemma}{Lemma}
\theoremstyle{definition}

\theoremstyle{remark}
\newtheorem*{remark}{Remark}
\newcommand{\mi}{\mathcal{I}}
\newcommand{\mj}{\mathcal{J}}
\newcommand{\mk}{\mathcal{K}}

\begin{document} \date{}
\allowdisplaybreaks
\IEEEoverridecommandlockouts

\title{Capacity of the Energy Harvesting Gaussian MAC}

\author{\IEEEauthorblockN{Huseyin A. Inan,  Dor Shaviv and Ayfer {\"{O}}zg{\"{u}}r \thanks{This work was presented in part at the 2016 IEEE International Symposium on Information Theory (ISIT)~\cite{isit16} and the 2016 14th International Symposium on Modeling and Optimization in Mobile, Ad Hoc and Wireless Networks (WiOpt)~\cite{wiopt}.} \thanks{This work was partly supported by a Robert Bosch Stanford Graduate Fellowship, by the National Science Foundation (NSF) under grant CCF-1618278 and the Center for Science of Information (CSoI), an NSF Science and Technology Center, under grant agreement CCF-0939370.} 
}

}
\maketitle

\begin{abstract}
We consider an energy harvesting multiple access channel (MAC) where the transmitters are powered by an exogenous stochastic energy harvesting process and equipped with finite batteries. We characterize the capacity region of this channel as $n$-letter mutual information rate and develop inner and outer bounds that differ by a constant gap. An interesting conclusion that emerges from our results is that the sum-capacity approaches that of a standard AWGN MAC (with only an average constraint on the transmitted power), as the number of users in the MAC becomes large.
\end{abstract}

\section{Introduction}
Energy harvesting wireless devices are expected to be one of the key enablers of the next exponential growth in wireless connectivity. By eliminating bulky batteries, decoupling node deployments from the power grid, and allowing wireless nodes to operate potentially forever in a maintenance-free manner, energy harvesting can enable massive deployments of wireless devices to connect objects and machines in the age of ``Internet of Everything'' (IoE). For example, in a home IoE application a house can be equipped with tens or even hundreds of wireless sensors and actuators that connect to the cloud through a central sink node. This common topology envisioned for IoE applications gives rise to an energy harvesting multiple-access channel, where a large number of energy harvesting wireless devices communicate to a central sink node which has access to traditional power.

Motivated by this observation, in this paper we study the information-theoretic capacity of such an energy-harvesting MAC where each transmitter is powered by an exogenous i.i.d. stochastic energy arrival  process and equipped with a finite battery. We allow for arbitrary battery sizes at the transmitters and arbitrary correlation between their energy arrival processes. We characterize the capacity of this channel as an $n$-letter mutual information rate. We then develop inner and outer bounds on this capacity region, which allow us to connect the capacity region of the energy harvesting MAC to the capacity region of the standard AWGN MAC under an average power constraint as well as to online power control policies over an energy harvesting MAC. %\cite{ozel1,mitran1,dorlatestarxiv}. 
Our approach follows and extends the approach developed in \cite{maincapacity}, \cite{maincapacityarticle} for point-to-point energy harvesting channels. An interesting conclusion that emerges from our results is that the sum-capacity approaches that of a standard AWGN MAC with an average power constraint, as the number of users in the MAC becomes large. While it has been known that an energy harvesting system can achieve the AWGN capacity in the limit when the battery size  becomes large \cite{ozel2}, it is interesting that the AWGN capacity can be also achieved asymptotically when the number of users becomes large. This is a natural limit for IoE networks which are envisioned to consist of massively large number of wireless devices.

The information-theoretic capacity of the point-to-point energy harvesting channel has been previously considered in  \cite{ozel2, mao1, dong1,  jog1,mao2, DorRBR, Ozeltimingchannel, Ozelzerobattery, maincapacity,maincapacityarticle}. Our work is most closely related to \cite{maincapacity}, \cite{maincapacityarticle} which develop $n$-letter expressions for the capacity and upper and lower bounds which differ by a constant gap. Power control and packet scheduling for the MAC channel has been considered in \cite{kaya2, mitran1, %wang1, 
yang2}. Building on \cite{dorarxiv}, in \cite{wiopt} we develop approximately optimal online power control policies, which as the current paper shows can be related to the information-theoretic capacity problem. \cite{rajesh1,ozel3} have previously considered the information theoretic capacity of the energy harvesting MAC when the battery size is infinite and zero respectively. To the best of our knowledge, the current paper is the first to treat the information-theoretic capacity of the energy harvesting MAC when transmitters have finite batteries.

\section{System Model}

We begin with the notation used throughout the paper. Let uppercase, lowercase, and calligraphic letters
denote random variables, specific realizations of RVs, and alphabets, respectively. For two jointly distributed RVs $(X, Y)$, $P_X$, $P_{X, Y}$, and
$P_{Y | X}$ are used to denote the marginal of $X$, the joint distribution of $(X, Y)$, and the conditional distribution of $Y$ given $X$ respectively. Let $\mathbb{E}[\cdot]$ denote
expectation. For a variable $X_i$, and for
$t \leq n$, let $X_{it}^n = (X_{it}, X_{i(t+1)}, \ldots, X_{in})$, and $X_{i}^n = X_{i1}^n$.   With  abuse  of
notation, we use superscript 2 to represent square, i.e., $X_{it}^2 = (X_{it})^2$. Furthermore, we use boldface letters to denote vectors when the length is clear from the context, infinite length mappings, and random processes where each definition is meticulously explained to avoid any confusion. All logarithms are to base 2 ($\ln$ will denote $\log$ to base $e$).

\begin{figure}[t!] 
  \begin{center} 
		\includegraphics[width=100mm]{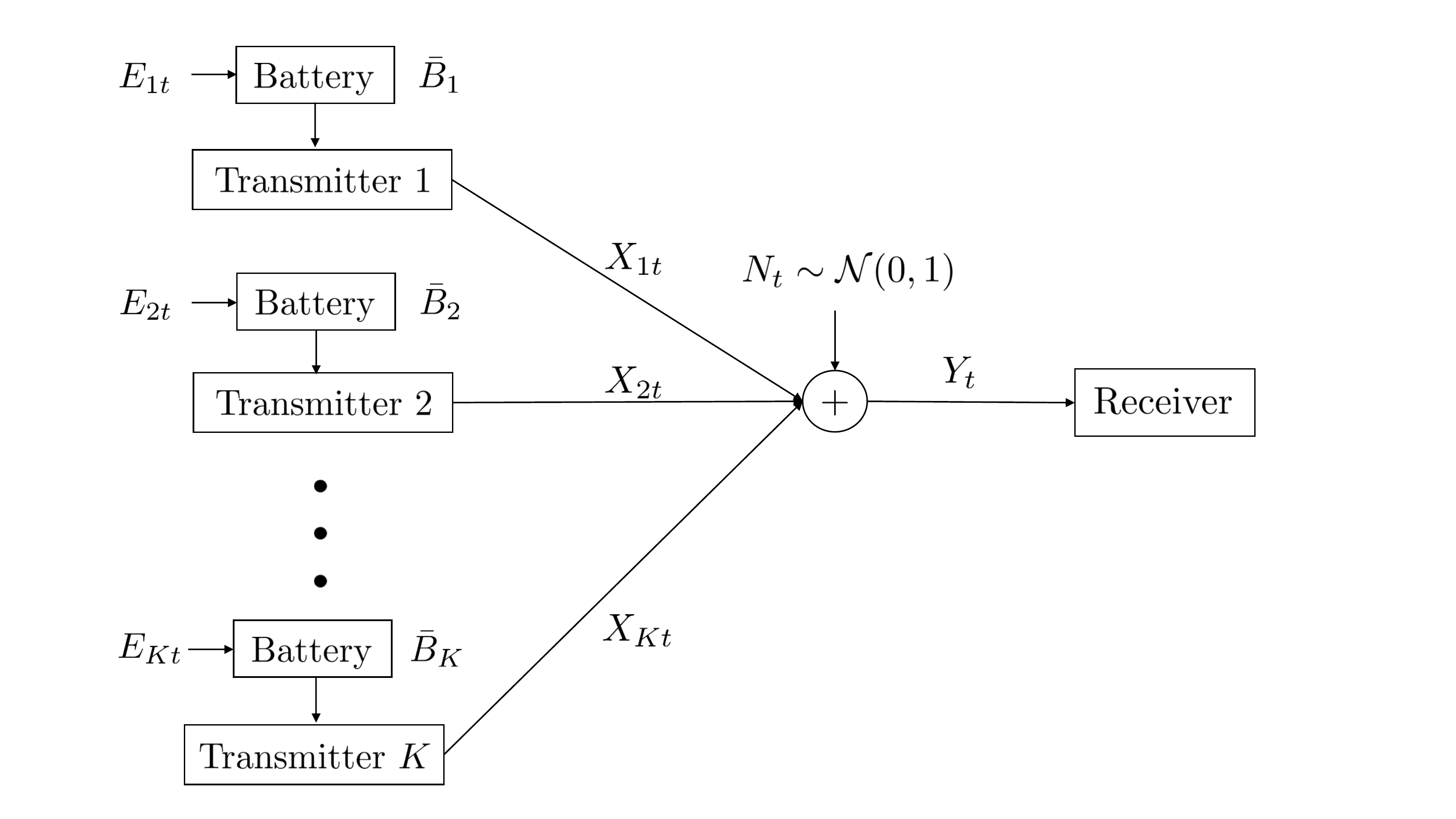} 	
		\caption{Energy harvesting AWGN MAC model.}
		\label{model}
  \end{center}
\end{figure}

We consider the $K$-sender discrete-time additive white Gaussian noise MAC (Gaussian MAC) model in Figure~\ref{model}. The channel output corresponding to the inputs $X_{1t}, \ldots, X_{Kt}$ transmitted by users $1, \ldots, K$ respectively at time t is
$
Y_t = X_{1t} + \ldots + X_{Kt} + N_t,
$
where $N_t \sim \mathcal{N}(0,1)$ and i.i.d. across time. The transmitters are equipped with rechargeable batteries with finite capacities $\bar{B}_1, \ldots, \bar{B}_K$, which are replenished by external energy arrival processes $E_{1t}, \ldots, E_{Kt}$ respectively. At each time $t$, the energy of the  symbol transmitted by  each transmitter is limited by the amount of energy available in its battery, i.e., for $i=1, \ldots, K,$
\begin{align}
& |X_{it}|^2 \leq {B}_{it}, \label{encos} \\
& {B}_{it} = \min \{   {B}_{i(t-1)} - |X_{i(t-1)}|^2 + {E}_{it}, \bar{B}_i \},
\label{encos2}
\end{align}
where  ${B}_{it}$ indicates the available energies at time $t$ in the battery of the corresponding transmitter. We assume that the energy arrival processes are i.i.d. across time and distributed according to some joint probability mass function $P_{E_1,\dots, E_K}(e_1,\dots,e_K)$ over the finite alphabets $\mathcal{E}_i$, such that
$E_i \geq 0$ and $\mathbb{E}[{E}_{it}] > 0$ for $i=1, \ldots, K$. Since the excess
energy cannot be used due to the battery constraints, without loss of generality we assume that $\mathcal{E}_i\subseteq [0, B_i]$. We point out that we do not make any assumptions regarding the joint distribution (e.g. independence or correlation between the energy arrival processes of the transmitters). It is shown in  \cite{maincapacityarticle} that the initial battery level does not change the capacity. This result in the point to point channel can be extended to the MAC model considered in this work to show that the initial battery levels do not change the capacity region. Therefore, without loss of generality we assume that the batteries are empty before the beginning of transmission, i.e., $B_{i0} = 0$ for $i = 1, \ldots, K$ and this is known to the transmitters and the receiver.

In this work, we will be interested in the capacity of this MAC under two different assumptions: 
\begin{itemize} 
\item[1)] The energy arrival processes are observed causally at the corresponding transmitters and not at the receiver; ${E}_{it}$ is
observed causally at the transmitter $i$ for $i=1, \ldots, K$. We denote the capacity region as $\mathcal{C}^{\text{Tx}}$ in this case. 
\item[2)] While each transmitter causally observes its energy arrival process as above, the receiver observes all processes. We denote the capacity region as $\mathcal{C}^{\text{TxRx}}$ in this case. 
\end{itemize} 
For the first case, we define a $((2^{nR_1}, \ldots, 2^{nR_K}),n)$ code consisting of encoding functions $f_{1t}, \ldots, f_{Kt}$ and a decoding function $g$:
\begin{align}
& f_{it} : \mathcal{W}_i \times \mathcal{E}_i^{t} \rightarrow \mathcal{X}_i, \hspace{10pt} t = 1, \ldots, n, \hspace{10pt} i = 1, \ldots, K,
\label{encodingfunctions} \\
& g : \mathcal{Y}^n \rightarrow  (\mathcal{W}_1 \times \ldots  \times \mathcal{W}_K),
\label{causalreceiver}
\end{align}
where $\mathcal{X}_i = \mathcal{Y} = \mathbb{R}$ for $i = 1, \ldots, K$ and $\mathcal{W}_i = \{1, 2, \ldots, 2^{nR_i}\}$. We assume that the message sequence $(W_1, \ldots, W_K)$ is uniformly distributed over $\mathcal{W}_1 \times \ldots \times \mathcal{W}_K$. To transmit the messages $w_i \in \mathcal{W}_i$, the transmitters set $X_{it} =  f_{it}(w_i,E_i^{t})$. Note that the battery states $B_{it}$ are deterministic functions of $(X_i^{t-1},E_i^{t})$, therefore also of $(w_i, E_i^{t})$. The functions $f_{it}$ must satisfy the energy constraint in \eqref{encos}, hence
\begin{align}
f_{it}(w_i, E_i^{t}) \leq B_{it}(w_i, E_i^{t})
\end{align}
for $i = 1, \ldots, K$.  The receiver sets $(\hat{W}_1, \ldots, \hat{W}_K) = g(Y^n)$ and the average probability of error for the \\
$((2^{nR_1}, \ldots, 2^{nR_K}),n)$ code
is defined as 
\begin{align}
P_e^{(n)}  =  \dfrac{1}{2^{n(R_1 + \ldots + R_K)}}  \sum \limits_{ 
\substack{ (w_1, \ldots, w_K) \in  \\ \mathcal{W}_1 \times \ldots \times \mathcal{W}_K }} \Pr \{ g(Y^n) \neq (w_1, \ldots, w_K) | 
(w_1, \ldots, w_K) \hspace{5pt} \text{sent}\}.
\end{align}
A rate tuple $(R_1, \ldots, R_K)$ is said to be achievable for the multiple access channel if there exists a sequence of $((2^{nR_1}, \ldots, 2^{nR_K}),n)$ codes with $P_e^{(n)}  \rightarrow 0$.
The capacity region $\mathcal{C}$ is defined as 
the closure of the set of achievable rate tuples $(R_1, \ldots, R_K)$.

When ${E}_{1t}, \ldots, {E}_{Kt}$ are observed at the receiver as well, we change \eqref{causalreceiver} to
$
g : \mathcal{Y}^n \times (\mathcal{E}_1^{n} \times \ldots \times \mathcal{E}_K^{n}) \rightarrow  (\mathcal{W}_1 \times \ldots \times \mathcal{W}_K).
$
%In this case the capacity region is denoted by $\mathcal{C}^{\text{TxRx}}$.

\section{Equivalent Channel Model}
Consider the case where the energy arrivals are observed causally at their corresponding transmitters. We utilize Shannon strategies (see \cite{shannonstr}, \cite{maincapacity} for a detailed discussion) to convert this channel into an equivalent channel with no state information at the transmitters but with a different input alphabet: the input of transmitter $i$ to the equivalent channel at time $t$ is a strategy letter $u_{it} : \mathcal{E}_i^{t} \rightarrow \mathcal{X}_i$ and the input alphabet for blocklength $n$ is of the form
\begin{align*}
\mathcal{U}_i^{n} = \{u_i^{n} \mid u_{it} : \mathcal{E}_i^{t} \rightarrow \mathcal{X}_i, \hspace{10pt} t = 1, \ldots, n \},
\end{align*}
for $i = 1, \ldots, K$. At time t, $X_{it} =
U_{it}({E}_i^{t})$ is transmitted over the original channel given the realization of ${E}_i^{t}$. The output
of the channel is the corresponding $Y_t \in \mathcal{Y}$ and the new channel is characterized by the following transition probabilities:
\begin{align}
 P_{Y^n | U_{1}^{n}, \ldots, U_{K}^{n}} (y^n | u_{1}^{n}, \ldots, u_{K}^{n}) &  = \sum \limits_{e_{1}^{n}, \ldots, e_{K}^{n}}   P_{E_{1}^{n}, \ldots, E_{K}^{n}}(e_{1}^{n}, \ldots, e_{K}^{n}) 
 P_{Y^n | X_{1}^{n}, \ldots, X_{K}^{n}} (y^n | u_{1}^{n}(e_{1}^{n}), \ldots, u_{K}^{n}(e_{K}^{n}))
\nn \\
& = \sum \limits_{e_{1}^{n}, \ldots, e_{K}^{n}}  \prod \limits_{t = 1}^{n} P_{E_{1}, \ldots, E_{K}}(e_{1t}, \ldots, e_{Kt}) 
P_{Y | X_{1}, \ldots, X_{K}} (y_t | u_{1t}(e_{1}^{t}), \ldots, u_{Kt}(e_{K}^{t})).
 \label{newchannel}
\end{align}
Note that there is no state in this new channel and the encoding functions \eqref{encodingfunctions} become $f_i : \mathcal{M}_i \rightarrow \mathcal{U}_{i}^{n} $ for $i = 1, \ldots, K$.
However, there are energy constraints restricting the admissible sets for $\mathcal{U}_{1}^{n}, \ldots, \mathcal{U}_{K}^{n}$ imposed by the constraints in our original energy harvesting channel. The energy constraints on our original energy harvesting channel imply that the admissible channel inputs $u_{i}^n$ should satisfy for every $e_{i}^n \in \mathcal{E}_i^n$:
\begin{align}
& |u_{it}(e_{i}^t)|^2 \leq {b}_{it}, \nn \\
& {b}_{it} = \min \{   {b}_{i(t-1)} - |u_{i(t-1)}(e_{i}^{t-1})|^2 + {e}_{it}, \bar{B}_i \}.
\label{channelconstraints}
\end{align}

We note that the capacity of this channel is the same as the original channel since coding strategies for one can
be immediately translated to the other. Therefore, when the energy arrival processes are observed only at the transmitters, we consider the equivalent channel introduced above and when the receiver also
observes the energy arrival processes, we consider the original
channel model.

\section{Channel Capacity}

Before stating the expressions for the capacity regions $\mathcal{C}^{\text{Tx}}$ and $\mathcal{C}^{\text{TxRx}}$,
we define the set of allowed input distributions on $\mathcal{U}_{i}^{n}$, $i = 1, \ldots, K$ for the equivalent channel as:
\begin{align}
\mathcal{P}_{i}^{n} = \big \lbrace & P_{{U}_{i}^{n}} \hspace{5pt} \text{s.t.} \hspace{5pt} \text{a.s.} \hspace{5pt} \text{for} \hspace{5pt} t = 1, \ldots, n \hspace{5pt} \text{and} \hspace{5pt} \forall e_{i}^{n} \in \mathcal{E}_{i}^{n} : \nn \\
& |{U}_{it}(e_{i}^{t})|^2 \leq B_{it}, \nn \\ & B_{it} = \min \lbrace B_{i(t-1)} - |{U}_{i(t-1)}(e_{i}^{t-1})|^2 + e_{it}, \bar{B}_i \rbrace
\big \rbrace.
\end{align} 
Note that we assign zero probability to any codeword that does not 
%satisfy \eqref{econst2} to 
obey the energy constraints.
For the case of energy arrival information available causally at the receiver as well, we use the notion of causal conditioning,
\begin{align}\label{eq:causal_conditioning}
P_{X_{i}^{n} || E_{i}^{n}}(x_{i}^{n} || e_{i}^{n}) = \prod \limits_{t = 1}^{n}
P_{X_{it} | X_{i}^{t-1},  E_{i}^{t}}(x_{it} | x_{i}^{t-1},  e_{i}^{t})
\end{align} 
for $i = 1, \ldots, K$. This differs from $P_{X^n|E^n}=\prod_{t=1}^{n}P_{X_t|X^{t-1},E^n}(x_t|x^{t-1},e^n)$ in that at time $t$ the dependence on $E^n$ is replaced by only the past and present $E^t$.
%This is more restrictive than $P_{X^n|E^n}$, in the sense that $X_t$ is independent of the future energy arrivals $E_{t+1}^{n}$ given $(X^{t-1},E^t)$.
Define
\begin{align}
\mathcal{Q}_{i}^{n} = \big \lbrace & P_{X_{i}^{n} || E_{i}^{n}}: P_{X_i^n|E_i^n}=P_{X_i^n\|E_i^n} \hspace{5pt} \text{s.t.} \hspace{5pt} \forall e_{i}^{n} \in \mathcal{E}_{i}^{n} \hspace{5pt}
\text{a.s.} \hspace{5pt} \text{for} \hspace{5pt} t = 1, \ldots, n: \nn \\
& |{X}_{it}|^2 \leq B_{it}, \nn \\ & B_{it} = \min \lbrace B_{i(t-1)} - |{X}_{i(t-1)}|^2 + e_{it}, \bar{B}_i \rbrace
\big \rbrace. 
\label{Qin} 
\end{align} 
This  imposes the additional constraint that $X_t$ must depend on $E_t$ in a causal manner, as defined in~\eqref{eq:causal_conditioning}. Note that $B_{it}$ is a function of $(X_i^{t-1},E_i^t)$, so $\mathcal{Q}_{i}^{n}$ is well-defined.

A sequence $\{ A_n \}$, $n = 1, 2, \ldots$, of regions in $\mathbb{R}^d$ is said to converge to a region $A$, written $A = \lim \limits_{n \rightarrow \infty} A_n$ if
\begin{align}
\limsup \limits_{n \rightarrow \infty} A_n = \liminf \limits_{n \rightarrow \infty} A_n = A,
\end{align}
where
\begin{align}
& \limsup \limits_{n \rightarrow \infty} A_n = \bigcap \limits_{n \geq 1}   \bigcup \limits_{m \geq n} A_m,   \\
& \liminf \limits_{n \rightarrow \infty} A_n =  \bigcup \limits_{n \geq 1}  \bigcap \limits_{m \geq n} A_m.
\end{align}
For a more detailed discussion on convergence of sets in finite dimensions see \cite{conset}.

With these definitions, we state the main theorem of the paper:
\begin{theorem}
\label{theoremcap} Let
$\mathbf{R} = (R_1, \ldots, R_K)$, $U_{\mi}^{n} = \{ U_i^n : i \in \mi \}$, $\mk = \{ 1, \ldots, K \}$ and $E_{\mk}^{n} = \{ E_1^n, \ldots, E_K^n \} $. The capacity regions of the energy harvesting MAC are given by 
\begin{align}
\mathcal{C}^{\text{Tx}} & = \lim  \limits_{n \rightarrow \infty} \mathcal{R}_n^{\text{Tx}} \\
\mathcal{C}^{\text{TxRx}} & = \lim  \limits_{n \rightarrow \infty} {\mathcal{R}}_n^{\text{TxRx}}
\end{align}
where 
\begin{align*}
\mathcal{R}_n^{\text{Tx}} = \bigcup 
\left\{ \mathbf{R} :
\begin{aligned}
& \sum \limits_{i \in \mi} R_i \leq \dfrac{1}{n} I(U_{\mi}^{n} ; Y^{n} | U_{\mi^c}^{n}) \\
& \text{for every} \hspace{5pt} \mi \subseteq \mk 
\end{aligned}
\right \}
\end{align*} 
and the union is over all $ P_{{U}_{\mk}^{n}}=\prod_i P_{{U}_{i}^{n}}$ s.t. $P_{{U}_{i}^{n}}\in \mathcal{P}_{i}^{n} \,\,\text{for}\,\, i = 1, \ldots, K$; and   
\begin{align*}
{\mathcal{R}}_n^{\text{TxRx}} =  \bigcup  
\left\{ \mathbf{R} :
\begin{aligned}
& \sum \limits_{i \in \mi} R_i \leq \dfrac{1}{n} I(X_{\mi}^{n} ; Y^{n}|X_{\mi^c}^{n}, E_{\mk}^{n}) \\
& \text{for every} \hspace{5pt} \mi \subseteq \mk
\end{aligned}
\right \}
\end{align*}
where the union is over all $P_{X_{\mk}^{n} | E_{\mk}^{n}}$ s.t. $P_{X_{\mk}^{n} | E_{\mk}^{n}}=\prod_i P_{X_{i}^{n} || E_{i}^{n}}$ and $P_{X_{i}^{n} || E_{i}^{n}} \in \mathcal{Q}_{i}^{n} \,\,\text{for}\,\,  i = 1, \ldots, K$.
\end{theorem}
The expressions for the capacity regions
$\mathcal{C}^{\text{Tx}}$ and $\mathcal{C}^{\text{TxRx}}$ are proved 
in Appendix \ref{sec:channel_capacity}.

\section{Capacity Bounds}

Due to the difficulty in evaluating these capacity regions, we next find outer and inner bound regions which are separated by a constant gap and relate to the resource allocation formulation of the energy harvesting communication problem extensively studied in the recent communication theory literature. Our approach follows and extends the approach developed in \cite{maincapacity}, \cite{maincapacityarticle} for point-to-point energy harvesting channels.

We start by stating a simple outer bound on the capacity region.
\begin{proposition}
The capacity region of the energy harvesting MAC  is bounded by
\begin{align}
\mathcal{C}^{\text{Tx}} \subseteq \mathcal{C}^{\text{TxRx}}
	\subseteq \overline{\mathcal{C}}
	\label{eq:TxRx_outer_bounds}
\end{align}
where
\[
\overline{\mathcal{C}}=\left\{
\mathbf{R}:
\begin{aligned}
& \sum \limits_{i \in \mi} R_i \leq \dfrac{1}{2} \log \left( 1 + \sum \limits_{i \in \mi} \mathbb{E}\left[ E_i \right]  \right)  \\
& \text{for every} \hspace{5pt} \mi \subseteq \mk
\end{aligned}
\right\}.
\]
\label{propObound}
\end{proposition}
The proof of this proposition follows from the fact that the average power at each transmitter can not exceed the average energy arrival rate $\mathbb{E}[E_i]$, $i=1, \ldots, K$. If this average power constraint was the only constraint imposed on the encoders the capacity region would be given by that of a standard AWGN MAC. Therefore this region provides an upper bound on the capacity region of the energy harvesting MAC. Note that it does not depend on the joint distribution of $E_1, \ldots, E_K$ but only on the expectations of the individual energy arrivals. We formally prove the proposition in Appendix \ref{sec:outer_bounds}.

Before stating our lower bounds, we introduce some terms and notations. An  online power control policy $\mathbf{g}_i$ for transmitter $i$ is a sequence of mappings $g_{it} : \mathcal{E}_i^t \rightarrow \mathbb{R}_{+}$ for $t = 1, \ldots$. An admissible policy is a policy that satisfies the energy constraints \eqref{encos} and \eqref{encos2}. In particular, the set of all admissible policies for transmitter $i$ is 
\begin{align}
\mathcal{G}_{i} = \bigg \lbrace & \mathbf{g}_i \hspace{5pt} | \hspace{5pt} 
%\text{s.t.}\hspace{5pt}  
\forall t,  \hspace{5pt}  \forall e^{t} \in \mathcal{E}_i^{t} : \nn \\* &
g_{it}(e^{t}) \leq b_{t}, \nn \\* & b_{t} = \min \lbrace b_{t-1} - g_{i(t-1)}(e^{t-1}) + e_{t}, \bar{B}_i \rbrace
\bigg \rbrace.
\label{admissibleonlinepolicy} 
\end{align} 

For any set of policies $\mathbf{g}_1, \ldots, \mathbf{g}_K$, define the $n$-horizon expected average throughput region achieved under a set of online policies  as 
the region $(R_1,\dots, R_K)$ which satisfies
$$
\sum_{i\in \mi} R_i\leq \frac{1}{n} \mathbb{E}\left[ \sum_{t=1}^n \frac{1}{2} \log\left(1+\sum_{i\in \mi} g_{it}(E_i^t)\right ) \right],
$$
for every $\mi \subseteq \{1, 2, \ldots, K\}$ where the expectation is taken over the distribution of the energy arrivals. We denote this region by $\mathcal{T}(g_{1}^{n}, g_{2}^{n}, \ldots, g_{K}^{n})$ and denote by $\mathcal{T}_{n}$ the union of the $n$-horizon
expected throughput regions over all possible online policies, i.e.,
\begin{align}
\mathcal{T}_{n} = \bigcup \limits_{g_{1}^{n}, g_{2}^{n}, \ldots, g_{K}^{n}} \mathcal{T}(g_{1}^{n}, g_{2}^{n}, \ldots, g_{K}^{n}).
\end{align}
The long-term average throughput region for the energy-harvesting MAC is defined as
\begin{align}
\mathcal{T} = \liminf \limits_{n \rightarrow \infty} \mathcal{T}_{n}.
\end{align}

The following theorem extends the result of \cite{dorarxiv} for the point-to-point case to the energy harvesting multiple access channel: 
\begin{theorem}
\label{theoremthroughput}
The long-term average throughput region of the energy harvesting MAC is bounded by
\begin{equation}
\underline{\mathcal{C}}(0.72)  \subseteq \mathcal{T} \subseteq \overline{\mathcal{C}},
\end{equation}
where  
\[
\underline{\mathcal{C}}(\gamma)=\left\{
\mathbf{R}:
\begin{aligned}
& \sum \limits_{i \in \mi} R_i \leq \dfrac{1}{2} \log \left( 1 + \sum \limits_{i \in \mi}  \mathbb{E}\left[ E_i \right]  \right)-\gamma  \\
& \text{for every} \hspace{5pt} \mi \subseteq \mk
\end{aligned}
\right\}.
\]
\end{theorem}

The theorem shows that when each of the energy-harvesting transmitters applies the fixed fraction online policy proposed in \cite{dorarxiv} for the point-to-point energy harvesting channel, i.e., at time $t$ the transmitter $i$ allocates energy 
\begin{equation}\label{fixedfrac}
g_{it} = q_i b_{it},
\end{equation}
where $q_i=\mathbb{E}[E_i]/\bar{B}_i$, then the long-term average throughput region achieved is within 0.72, the same gap as in the point-to-point case in \cite{dorarxiv}, of the AWGN capacity region in Proposition~\ref{propObound} 
which is also an outer region for the capacity region of the energy harvesting MAC. Note that in order to apply this strategy, the users  only need to know the mean of their own energy arrival process and not the exact distribution. Moreover, they do not need to know anything about the distribution of the energy arrival processes at the other users, how it correlates with their own process or their battery capacity. The strategy is universal in the sense that it performs well under any joint distribution of the harvesting processes.

The theorem is proved in Appendix~\ref{sec:powercontrolMAC}. Note that it holds regardless of the correlation structure of the energy harvesting processes
at different users.

We next connect the online power control problem to the information theoretic capacity region. For any set of online policies $\mathbf{g}_1, \ldots, \mathbf{g}_K$,
let $\mathbf{g}_{\mi} = \{ \mathbf{g}_i : i \in \mi \}$ for $ \mi \subseteq \{ 1, \ldots, K \}$. Define
\begin{align*}
\mathscr{T}(\mathbf{g}_{\mi})= \liminf  \limits_{n\to\infty}\frac{1}{n}\mathbb{E}\left[\sum_{t=1}^{n}
	\frac{1}{2}\log \left(  1 + \sum \limits_{i \in \mi} g_{it}(E_i^t) \right) \right].
\end{align*}
This quantity can be interpreted as the sum long-term average throughput for the subset $\mi$. We also define the entropy rate of the set of policies $\mathbf{g}_1, \ldots, \mathbf{g}_K$ as
$H(\mathbf{g}_1(\mathbf{E}_1), \ldots, \mathbf{g}_K(\mathbf{E}_K) )=\limsup \limits_{n\to\infty}\frac{1}{n}H\left(  {g}_1^n(E_1^n), \ldots, {g}_K^n(E_K^n) \right) $. Note that ${g}_{it}(E_i^t)$ is a discrete random variable since it is a deterministic function of a discrete random vector $E_i^t$. Therefore, the discrete entropy $H(\cdot)$ is well-defined. We further point out that
our definition of entropy rate enables us to take into account the case when the limit may not exist. This provides a general framework for our results.

We next state our lower bounds for the information theoretic capacity in two steps.
\begin{theorem}
\label{theorem2}
For any admissible $\mathbf{g}_1\in\mathcal{G}_1, \ldots, \mathbf{g}_K\in\mathcal{G}_K$, the capacity region of the energy harvesting MAC is bounded by
\begin{align}
\underline{\mathcal{C}}^{\text{TxRx}}(\mathbf{g}_1, \ldots, \mathbf{g}_K)
&\subseteq
\mathcal{C}^{\text{TxRx}}
\nn \\
\underline{\mathcal{C}}^{\text{Tx}}(\mathbf{g}_1, \ldots, \mathbf{g}_K)
	&\subseteq\mathcal{C}^{\text{Tx}}
	\label{eq:TxRx_bounds_g}
\end{align}
where
\[
\underline{\mathcal{C}}^{\text{TxRx}}(\mathbf{g}_1, \ldots, \mathbf{g}_K)=
\left\{ \mathbf{R}:
\begin{aligned}
& \sum \limits_{i \in \mathcal{I}} R_i \leq \mathscr{T} \left(  \mathbf{g}_{\mathcal{I}} \right)  -1.05
\\
& \text{for every} \hspace{5pt} \mathcal{I} \subseteq \mathcal{K}
\end{aligned}	
\right\}
\]
and 
\[ 
\underline{\mathcal{C}}^{\text{Tx}}(\mathbf{g}_1, \ldots, \mathbf{g}_K)=
\left\{ \mathbf{R}:
\begin{aligned}
& \sum \limits_{i \in \mathcal{I}} R_i \leq\mathscr{T}(\mathbf{g}_{\mathcal{I}})-H(\mathbf{g}_1(\mathbf{E}_1), \ldots, \mathbf{g}_K(\mathbf{E}_K) )-1.05\\
& \text{for every} \hspace{5pt} \mathcal{I} \subseteq \mathcal{K}
\end{aligned}	
\right\}.
\]
\end{theorem}
The above theorem states that any sequence of online power control policies can be used to derive inner bounds on the capacity of the energy harvesting MAC. To maximize the inner bound for $\mathcal{C}^{\text{TxRx}}$ we would need an online power control policy that maximizes the long-term average throughput. In particular, 
by combining the results of Theorems~\ref{theoremthroughput} and \ref{theorem2}, we can show the following proposition.
\begin{proposition} For the energy harvesting MAC with side information at the receiver, we have
$$
\underline{\mathcal{C}}(1.77)
	\subseteq\mathcal{C}^{\text{TxRx}}
	\nn.
$$
\end{proposition}
We skip the proof of the proposition. The proposition simply follows by inserting the lower bounds achieved for $\mathscr{T} \left(  \mathbf{g}_{\mathcal{I}} \right)$ for all $\mi \subseteq \mk$ in the proof of Theorem \ref{theoremthroughput} to the expression
$\underline{\mathcal{C}}^{\text{TxRx}}(\mathbf{g}_1, \ldots, \mathbf{g}_K)$.

To maximize the inner bound for $\mathcal{C}^{\text{Tx}}$ in Theorem~\ref{theorem2}, we need to also take into account the entropy rate of the power control policies. This is because there is a rate penalty, given by the entropy rate of the power allocation process, for enabling the receiver to be able to track the power allocations chosen by the transmitters. The fixed fraction policy in \eqref{fixedfrac} used to prove Theorem~\ref{theoremthroughput} can have an entropy rate that can be arbitrarily large depending on the distribution of the energy arrivals. For the case when the reciever does not have energy arrival information, we instead use the online power control policy developed in \cite[Theorem 3]{maincapacityarticle} for the point-to-point energy harvesting channel, whose entropy rate can be bounded by $1$ bits/channel use indepedent of the distribution of the energy arrivals. By applying this strategy independently at all transmitters, we  obtain the following theorem:

\begin{theorem}
\label{theorem3}
The capacity region of the energy harvesting MAC without energy arrival information at the receiver is bounded by 
\begin{equation}
	\underline{\mathcal{C}}(2.85 + K)
	\subseteq\mathcal{C}^{\text{Tx}}.
	\label{eq:TxRx_bounds_gamma}
\end{equation}
\end{theorem}

When contrasted with the approximation results for the point-to-point energy harvesting channel  developed in \cite{maincapacityarticle}, it is interesting to note that in the case where the transmitters and the receiver have information regarding the energy harvesting process, the gap to the standard AWGN capacity is exactly the same. While it is natural that the gap for the individual rate constraints is the same as in the point-to-point case, it is interesting to note that we can maintain the same gap for the sum rate constraints. In the next section, we build on this observation to argue that we can achieve  the AWGN capacity  with diminishing gap in the limit when the number of users in the MAC  becomes large. For the case when only the transmitters observe the energy arrival process, the gap increases from $3.85$ in the point-to-point case to $2.85 + K$ bits/channel use. This is because the entropy rate of the power control policy developed in \cite[Theorem 3]{maincapacityarticle}, that allows us to obtain Theorem \ref{theorem3} from Theorem \ref{theorem2}, can be trivially bounded as 
\begin{equation}\label{eq:entropy}
H(\mathbf{g}(\mathbf{E})) \leq 1,
\end{equation}
for any energy arrival process $E_t$. When the same strategy is used at all transmitters in the MAC we can bound $H(\mathbf{g}(\mathbf{E}_1), \ldots,  \mathbf{g}(\mathbf{E}_K)) \leq K$. This however increases the gap by $K$ bit/channel use with respect to the point-to-point case.

The inner bounds in the last two theorems are proved in Appendix \ref{sec:lower_bounds}.

\section{Sum-Capacity of the Energy Harvesting Gaussian MAC}

In this section, we will be interested in the sum-capacity of the energy harvesting Gaussian MAC which we denote by $C_{\text{sum}}$. We begin with the case where the energy arrival processes are observed at the corresponding transmitters and the receiver.

\subsection{Energy Arrival Information at the Transmitters and the Receiver}\label{sec:1}

In the case when there is energy arrival information at the corresponding transmitters and the receiver, we prove the following proposition in Appendix \ref{sec:kuserTxRx}, which shows that the sum-capacity is dictated by the sum rate constraint in the capacity region:
\begin{proposition} \label{propsym}
When the receiver has side information  regarding the energy arrival process at the transmitter, the sum-capacity is given by
\begin{align*}
C_{\text{sum}}^{\text{TxRx}} = \lim \limits_{n \rightarrow \infty} \sup \limits_{\substack{ P_{X_{\mk}^{n} | E_{\mk}^{n}}=\prod_i P_{X_{i}^{n} || E_{i}^{n}}\\  P_{X_{i}^{n} || E_{i}^{n}} \in \mathcal{Q}_{i}^{n}  \\ i = 1, \ldots, K}}  \dfrac{1}{n} I(X_{\mk}^{n} ;Y^{n}|E_{\mk}^{n}).
\end{align*}
\end{proposition}

Proposition \ref{propObound} shows that the region $\overline{\mathcal{C}}$ is an outer bound for the capacity region $\mathcal{C}^{\text{TxRx}}$. By taking $\mi = \mk$, this allows us to immediately obtain the following upper bound on $C_{\text{sum}}^{\text{TxRx}}$ :
\begin{align*}
C_{\text{sum}}^{\text{TxRx}} \leq \dfrac{1}{2} \log \left( 1 + K \mathbb{E}\left[ E \right]  \right).
\end{align*}
By following similar  steps as in the proof of Theorem \ref{theoremthroughput} and Theorem \ref{theorem2}, we can obtain the following lower bound on $C_{\text{sum}}^{\text{TxRx}}$:
\begin{align*}
\dfrac{1}{2} \log \left( 1 + K \mathbb{E}\left[ E \right]  \right)- 1.77\leq C_{\text{sum}}^{\text{TxRx}}.
\end{align*} 

We combine these results and formally prove the following proposition in Appendix \ref{sec:kuserTxRx}.
\begin{proposition} \label{KuserTXRXbounds}
When the receiver has side information  regarding the energy arrival process at the transmitter, the  sum-capacity is
bounded by
\begin{align}\label{eq:symC1}
\dfrac{1}{2} \log \left( 1 + K \mathbb{E}\left[ E \right]  \right)- 1.77\leq C_{\text{sum}}^{\text{TxRx}} \leq  \dfrac{1}{2} \log \left( 1 + K \mathbb{E}\left[ E \right]  \right).
\end{align} 
\end{proposition}

We note that $C_{\text{sum}}^{\text{TxRx}}$ increases with the number of users  therefore the constant gap in the above approximation becomes negligible. While it has been known that an energy harvesting system can achieve the AWGN capacity in the limit when the battery size  becomes large \cite{ozel2}, it is interesting to note that the AWGN capacity can be also achieved asymptotically when the number of users becomes large. This can be a more natural limit for energy harvesting networks, for example in the context of IoT, which are expected to consist of a massively large number of tiny devices. 

Note that this result holds for any correlation of the energy arrivals at the transmitters, including independent or fully correlated arrivals. A priori, it is not clear if correlation of the energy arrivals would increase or decrease the capacity region. On one hand, correlation of the energy arrivals can allow transmitters to guess each other's energy states and adapt accordingly. On the other hand, if the high and low energy states of the transmitters correlate, intuitively this can  increase ``clashes'' over the MAC. While Proposition~\ref{KuserTXRXbounds} does not resolve this question of whether correlation helps or  hurts, it says that when the receiver has side information regarding the energy arrivals, the impact of correlation on capacity is limited by a constant. The discussion in the next section suggests that when the receiver does not know the energy arrivals, correlation can  lead to a larger capacity.

\subsection{Energy Arrival Information at the Transmitters Only}

In the case when the energy arrival processes are observed causally at the corresponding transmitters and not at the receiver, we similarly show in Appendix \ref{sec:kuserTxRx} that the sum-capacity is dictated by the sum rate constraint in the capacity region:
\begin{proposition} \label{propsymonlyTX}
The sum-capacity when the energy arrival information is
present at only the transmitters but not the receiver is given by
\begin{align*}
C_{\text{sum}}^{\text{Tx}} = \lim \limits_{n \rightarrow \infty} \sup \limits_{\substack{{P_{{U}_{\mk}^{n}}=\prod_i P_{{U}_{i}}^{n}} \\ P_{{U}_{i}^{n}}\in \mathcal{P}_{i}^{n}  \\ i = 1, \ldots, K  }}  \dfrac{1}{n} I(U_{\mk}^{n};Y^{n}).
\end{align*}
\end{proposition}
We note that following a similar approach as in Proposition \ref{KuserTXRXbounds} would lead us to a gap between the upper and lower bound for $C_{\text{sum}}^{\text{Tx}}$ that can increase linearly in $K$ because of the extra entropy rate term in the approximation gap. However, if the energy arrival processes are correlated with each other, the entropy rate term can be smaller than order $K$. Therefore, in this section,  we focus on a special case of the energy harvesting MAC and assume that the energy arrivals at different transmitters are fully correlated, i.e. $E_1=E_2=\dots=E_K=E$ and the transmitters are equipped
with batteries of the same size $\bar{B}$. This can model the scenario where all the transmitters harvest energy from the same physical process. 

The assumption of the energy arrivals being fully correlated and the battery sizes being same allows us to conclude that when all the transmitters use  the online policy developed in \cite[Theorem 3]{maincapacityarticle}
\begin{equation}\label{eq:entropy2}
H(\mathbf{g}(\mathbf{E}_1), \ldots, \mathbf{g}(\mathbf{E}_K))=H(\mathbf{g}(\mathbf{E})) \leq 1.
\end{equation}
This implies that the gap in the lower bound in Proposition~\ref{KuserTXRXbounds} increases only to $3.85$ bit/channel use in the case when the receiver does not have side information.  The formal proof of the following proposition is left to Appendix
\ref{sec:kuserTxRx}.
\begin{proposition} \label{KuserTXbounds}
When the energy arrivals at different transmitters with batteries of the same size $\bar{B}$ are fully correlated, i.e. $E_1=E_2=\dots=E_K=E$ and the energy arrival information is
present at only the transmitters but not the receiver, the sum-capacity can be bounded as
\begin{equation}
\dfrac{1}{2} \log\left( 1 +  K \mathbb{E}\left[ E \right]  \right) -  3.85  
\leq C_{\text{sum}}^{\text{Tx}} \leq  \dfrac{1}{2} \log \left( 1 + K \mathbb{E}\left[ E \right]  \right). 
\label{eq:last} 
\end{equation}
\end{proposition} 

Again the gap to the AWGN capacity becomes negligible when the number of users becomes large.  Note that the reason why we restrict attention to fully correlated energy arrivals in this section is to be able to upper bound the total entropy rate of the energy arrivals by $1$ as in \eqref{eq:entropy2}. When the energy arrivals are independent, the entropy rate in \eqref{eq:entropy2} can grow linearly in $K$ which would lead to a linear gap in the approximation in \eqref{eq:last}. This implies that, at least from the perspective of our approximation results, having independent energy arrivals at different users can be harmful when the receiver does not have side information regarding the energy arrival process as there seems to be a rate penalty associated with learning the energy arrivals at the receiver.

\section{Conclusion}

In this paper, we studied the capacity region of the energy-harvesting MAC, where transmitters are powered by an external energy arrival process. We characterized the capacity region as an $n$-letter mutual information rate and derived inner and outer bounds on the capacity region which differ by a constant gap. An interesting consequence of our results is that when the energy arrivals at different transmitters with batteries of the same size are identical (but still random and i.i.d. over time), the sum-capacity of the channel approaches to the standard AWGN capacity (under an average power constraint) in the limit when the number of transmitters becomes large. It would be interesting to extend the framework of this paper to other multi-user settings.

\bibliographystyle{IEEEbib}
\bibliography{bibfile}

\appendices
\section{Channel Capacity}
\label{sec:channel_capacity}

\subsection{Proof of Theorem \ref{theoremcap}}

We first derive the capacity region $\mathcal{C}^{\text{Tx}}$ and we then continue with $\mathcal{C}^{\text{TxRx}}$.

\subsubsection{Energy Arrival Information at the Transmitters Only} 
We begin with proof of achievability for the capacity region $\mathcal{C}^{\text{Tx}}$. We will follow a similar approach introduced in \cite{maincapacityarticle} for the point-to-point energy harvesting channel. The communication will occur in $k$ blocks where each block contains the codewords of length $n$
of the transmitters from the corresponding set $\mathcal{U}_{i}^{n}$.
Therefore, we fix $P_{{U}_{i}^{n}} \in \mathcal{P}_{i}^{n}$ and for each message $w_i$, we generate $k$ random codewords independently $\mathbf{u}_{ij} \sim P_{{U}_{i}^{n}}$, for $i = 1, \ldots, K$ and $j = 1, \ldots, k$. The chosen messages will be transmitted within each block where the size of a block is $n$. 

We note that the random codewords are generated by distributions that assume an empty battery in the beginning and the battery levels in the beginning of each block will be at least 0. We also note that the set of allowed distributions are 
constructed such that any codeword that does not obey the energy constraints is assigned with zero probability. Therefore, the energy constraints are satisfied with this strategy. We further point out that the blocks are decoupled in the sense that they are independent of each other.

Denote the channel output during  block $j$
by $\mathbf{y}_j = y_{(j-1)n+1}^{(j-1)n+n}$ and the energy arrivals during the transmission of the codeword $ \mathbf{u}_{ij} $ by $\mathbf{e}_{ij} = {e}_{i, (j-1)n+1}^{(j-1)n+n}$. The receiver observes and uses $\mathbf{y}^{k} = (\mathbf{y}_1, \ldots, \mathbf{y}_k)$ for decoding, by applying standard jointly typical decoding with $\mathbf{u}_1^k, \ldots, \mathbf{u}_K^k$. We obtain the following channel transition probability from $\mathbf{U}_{1}^{k}, \ldots,  \mathbf{U}_{K}^{k}$ to $\mathbf{Y}^{k}$:
\begin{align}
 P_{\vY | \vUo, \ldots,  \vUt} (\vy | \mathbf{u}_1^k, \ldots, \mathbf{u}_K^k)  & = \sum \limits_{e_{1}^{kn}, \ldots, e_{K}^{kn}}  P_{E_{1}^{kn}, \ldots, E_{K}^{kn}} \left(   e_{1}^{kn}, \ldots, e_{K}^{kn} \right)
P_{Y^{kn} | X_{1}^{kn}, \ldots, X_{K}^{kn}} 
\left( \vy | \mathbf{u}_1^k(\mathbf{e}_1^k), \ldots, \mathbf{u}_K^k(\mathbf{e}_K^k)  \right)
\nn \\ & = 
\sum \limits_{e_{1}^{kn}, \ldots, e_{K}^{kn}} \prod \limits_{j = 1}^{k} 
P_{E_{1}^{n}, \ldots, E_{K}^{n}} \left(   \mathbf{e}_{1j}, \ldots, \mathbf{e}_{Kj} \right)
P_{Y^{n} | X_{1}^{n}, \ldots, X_{K}^{n}}\left(
\mathbf{y}_j | \mathbf{u}_{1j}(\mathbf{e}_{1j}), \ldots, \mathbf{u}_{Kj}(\mathbf{e}_{Kj}) \right) 
\nn \\ & = 
 \prod \limits_{j = 1}^{k} \sum \limits_{\mathbf{e}_{1j}, \ldots, \mathbf{e}_{Kj}}
P_{E_{1}^{n}, \ldots, E_{K}^{n}} \left(   \mathbf{e}_{1j}, \ldots, \mathbf{e}_{Kj} \right)
P_{Y^{n} | X_{1}^{n}, \ldots, X_{K}^{n}}\left(
\mathbf{y}_j | \mathbf{u}_{1j}(\mathbf{e}_{1j}), \ldots, \mathbf{u}_{Kj}(\mathbf{e}_{Kj}) \right).
\end{align}
Note that since $(E_{1t}, \ldots, E_{Kt})$ is i.i.d. across time, $P_{E_{1}^{n}, \ldots, E_{K}^{n}}$ do not depend on $j$,
so this is a memoryless channel with transition probability
\begin{align}
P_{\mathbf{Y} | \mathbf{U}_1, \ldots, \mathbf{U}_K}(\mathbf{y} | \mathbf{u}_1,
\ldots, \mathbf{u}_K)  & = \sum \limits_{{e}_{1}^{n}, \ldots, {e}_{K}^{n}} P_{E_{1}^{n}, \ldots, E_{K}^{n}} ({e}_{1}^{n}, \ldots, {e}_{K}^{n}) 
 P_{Y^{n} | X_{1}^{n}, \ldots, X_{K}^{n}} \left( \mathbf{y} | \mathbf{u}_{1}(e_{1}^{n}),   \ldots, \mathbf{u}_{K}(e_{K}^{n}) \right) \nn \\ & =  
 P_{Y^{n} | U_{1}^{n}, \ldots, U_{K}^{n}} \left( \mathbf{y} | 
 \mathbf{u}_{1}, \ldots, \mathbf{u}_{K}  \right),
\end{align}
where the last step is from \eqref{newchannel}. Note that $\mathbf{Y} = Y^{n}$ is the output of the channel from $U_{1}^{n}, \ldots, U_{K}^{n}$ to $Y^{n}$.
Taking $k \rightarrow \infty$, we get by standard joint typicality arguments that the following region is achievable
\begin{align}
\left\{ \mathbf{R} :
\begin{aligned}
& \sum \limits_{i \in \mi} R_i \leq \dfrac{1}{n} I(U_{\mi}^{n} ; Y^{n} | U_{\mi^c}^{n}) \\
& \text{for every} \hspace{5pt} \mi \subseteq \mk 
\end{aligned}
\right \},
\end{align}
where $U_\mi^{n} = \{ U_i^n : i \in \mi \}$, $\mathbf{R} = (R_1, \ldots, R_K)$ and $\mk = \{ 1, \ldots, K \}$. Since the input distributions were arbitrary, we have the following achievable region
\begin{align}
\mathcal{R}_n^{\text{Tx}} = \bigcup 
\left\{ \mathbf{R} :
\begin{aligned}
& \sum \limits_{i \in \mi} R_i \leq \dfrac{1}{n} I(U_{\mi}^{n} ; Y^{n} | U_{\mi^c}^{n}) \\
& \text{for every} \hspace{5pt} \mi \subseteq \mk 
\end{aligned}
\right \},
\label{Rn}
\end{align} 
where the union is over all $ P_{{U}_{\mk}^{n}}=\prod_i P_{{U}_{i}^{n}}$ s.t. $P_{{U}_{i}^{n}}\in \mathcal{P}_{i}^{n} \,\,\text{for}\,\, i = 1, \ldots, K$. Consequently,  the region 
$\cup_n \mathcal{R}_n^{\text{Tx}} $ is achievable, i.e.,
\begin{align}
 \bigcup_n \mathcal{R}_n^{\text{Tx}} \subseteq \mathcal{C}^{\text{Tx}}.
\end{align}
By definition, we have 
\begin{align}
 \limsup \limits_{n \rightarrow \infty} \mathcal{R}_n^{\text{Tx}} \subseteq   \bigcup_n \mathcal{R}_n^{\text{Tx}}  \subseteq \mathcal{C}^{\text{Tx}}.
\end{align}

For the converse part, consider any sequence of $((2^{nR_1}, \ldots, 2^{nR_K}),n)$ code with $P_e^{(n)} \rightarrow 0$. Fix $n$ and consider the given code of block length $n$. The joint distribution on $\mathcal{W}_1 \times \ldots \times \mathcal{W}_K \times \mathcal{U}_1^n \times \ldots \times \mathcal{U}_K^n \times \mathcal{Y}^n$ is well defined. Due to the channel constraints \eqref{channelconstraints}, the induced distributions $P_{{U}_{1}^{n}}, \ldots, P_{{U}_{K}^{n}}$ will satisfy $P_{{U}_{i}^{n}} \in \mathcal{P}_i^n$.
By Fano's inequality,
\begin{align*}
H(W_1, \ldots, W_K|Y^{n}) & \leq H(P_e^{(n)}) + P_e^{(n)} n (R_1 + \ldots + R_K) \\ & = n \epsilon_n, 
\end{align*}
where it is clear that $\epsilon_n \rightarrow 0$ as $P_e^{(n)} \rightarrow 0$. For any $ \mi \subseteq \{ 1, \ldots, K \} $, we have
\begin{align*}
n \sum \limits_{i \in \mi} R_i & = H(W_\mi)
\nn \\ & = H(W_\mi | U_{\mi^c}^n) \nn \\ & =
I(W_\mi ; Y^n | U_{\mi^c}^n) + H(W_\mi | Y^n, U_{\mi^c}^n).
\end{align*}
Since
\begin{align*}
H(W_\mi | Y^n, U_{\mi^c}^n) & \leq H(W_\mi | Y^n) \\
& \leq H(W_1, \ldots, W_K | Y^n) \nn
\\ & \leq   n \epsilon_n,
\end{align*}
and
\begin{align*}
I(W_\mi ; Y^n | U_{\mi^c}^n) & \leq I(W_\mi, U_{\mi}^n ; Y^n | U_{\mi^c}^n) \nn \\* & =
I(U_{\mi}^n ; Y^n | U_{\mi^c}^n) + I(W_{\mi} ; Y^n | U_{1}^n, \ldots, U_{K}^n)
\nn \\* & =
I(U_{\mi}^n ; Y^n | U_{\mi^c}^n),
\end{align*}
where the last equality is due to the Markov Chain $(W_1, \ldots, W_K) - (U_{1}^{n}, \ldots, U_{K}^{n}) - Y^n$, we have
\begin{align*}
\sum \limits_{i \in \mi} R_i \leq   \dfrac{1}{n} I(U_{\mi}^n ; Y^n | U_{\mi^c}^n) +  \epsilon_n.
\end{align*}
We conclude that the achievable rate tuple $(R_1, \ldots, R_K)$ satisfies
\begin{align}
\sum \limits_{i \in \mi} R_i \leq   \dfrac{1}{n} I(U_{\mi}^n ; Y^n | U_{\mi^c}^n) +  \epsilon_n
\label{converseee}
\end{align}
for every $ \mi \subseteq \{ 1, \ldots, K \} $ and every $n \geq 1$.

We define the following region
\begin{align}
\bar{\mathcal{R}}_n^{\text{Tx}} = \bigcup 
\left\{ \mathbf{R} :
\begin{aligned}
& \sum \limits_{i \in \mi} R_i \leq \dfrac{1}{n} I(U_{\mi}^{n} ; Y^{n} | U_{\mi^c}^{n}) + \epsilon_n \\
& \text{for every} \hspace{5pt} \mi \subseteq \mk
\end{aligned}
\right \},
\label{barRn}
\end{align}
where the union is over all $ P_{{U}_{\mk}^{n}}=\prod_i P_{{U}_{i}^{n}}$ s.t. $P_{{U}_{i}^{n}}\in \mathcal{P}_{i}^{n} \,\,\text{for}\,\, i = 1, \ldots, K$. Since for each $n$, there exists a distribution $P_{{U}_{1}^{n}} \times \ldots \times P_{{U}_{K}^{n}}$ ($P_{{U}_{i}^{n}} \in \mathcal{P}_i^n$) such that an achievable rate tuple $(R_1, \ldots, R_K)$ satisfies \eqref{converseee}, it follows that
\begin{align}
\mathcal{C}^{\text{Tx}} \subseteq \bigcap \limits_n 
\bar{\mathcal{R}}_n^{\text{Tx}}.
\end{align}
By definition, 
\begin{align}
\bigcap \limits_n 
\bar{\mathcal{R}}_n^{\text{Tx}} \subseteq \liminf \limits_{n \rightarrow \infty} \bar{\mathcal{R}}_n^{\text{Tx}},
\end{align}
therefore, we conclude that
\begin{align}
\limsup \limits_{n \rightarrow \infty} \mathcal{R}_n^{\text{Tx}} \subseteq      \mathcal{C}^{\text{Tx}} \subseteq  \liminf \limits_{n \rightarrow \infty} \bar{\mathcal{R}}_n^{\text{Tx}}.
\end{align}
We will prove that $\liminf  \limits_{n \rightarrow \infty} \mathcal{R}_n^{\text{Tx}} = \liminf  \limits_{n \rightarrow \infty} \bar{\mathcal{R}}_n^{\text{Tx}}$  to conclude that $\mathcal{C}^{\text{Tx}} = \lim \limits_{n \rightarrow \infty} \mathcal{R}_n^{\text{Tx}}$.

The Hausdorff distance between two sets $A$ and $B$, denoted by $h(A, B)$, is defined as follows
\begin{align*}
h(A, B) = \max \left\lbrace   \sup \limits_{\mathbf{x} \in A}  d(\mathbf{x}, B), \sup\limits_{\mathbf{y} \in B}  d(\mathbf{y}, A)  \right\rbrace 
\end{align*}
where the distance between a set $B$ and a point $\mathbf{x} \in A$ is given by
\begin{align}
d(\mathbf{x}, B) = \inf \limits_{\mathbf{y} \in B} || \mathbf{x} - \mathbf{y} ||.
\end{align}
We now state Lemma 8 of \cite{permuter1} which we will utilize in our proof:
\begin{lemma}
If $\lim \limits_{n \rightarrow \infty } h(A_n, B_n) = 0$, then $\limsup \limits_{n \rightarrow \infty } A_n = \limsup \limits_{n \rightarrow \infty } B_n$,  and  $\liminf \limits_{n \rightarrow \infty } A_n = \liminf \limits_{n \rightarrow \infty } B_n$.
\end{lemma}

Note that the results hold regardless of the distance measure chosen. Hence, we take $l_1$ distance, i.e., $l_1(\mathbf{x} , \mathbf{y}) = || \mathbf{x} - \mathbf{y} ||_1$ in our derivations. From \eqref{barRn} and \eqref{Rn}, we note that $\mathcal{R}_n^{\text{Tx}} \subseteq  \bar{\mathcal{R}}_n^{\text{Tx}}$ hence $d(\mathbf{x}, \bar{\mathcal{R}}_n^{\text{Tx}}) = 0$ for any $\mathbf{x} \in \mathcal{R}_n^{\text{Tx}}$, therefore
\begin{align}
\sup \limits_{\mathbf{x} \in \mathcal{R}_n^{\text{Tx}}}  d(\mathbf{x}, \bar{\mathcal{R}}_n^{\text{Tx}}) = 0.
\end{align}
We further note that from the inequalities in  \eqref{barRn} and \eqref{Rn}, for any $\mathbf{y} \in \bar{\mathcal{R}}_n^{\text{Tx}}$, if we take the point $\mathbf{x} = (\mathbf{y} - \mathbf{1} \epsilon_n)^+$, where $\mathbf{1} = (1, \ldots, 1)$ and $\mathbf{y}^+ = (\max(y_1, 0), \ldots, \max(y_K, 0))$,
we then observe that 
$ \mathbf{x} \in \mathcal{R}_n^{\text{Tx}} $
and therefore it follows that 
\begin{align}
\sup \limits_{\mathbf{y} \in \bar{\mathcal{R}}_n^{\text{Tx}}}   d(\mathbf{y}, \mathcal{R}_n^{\text{Tx}}) \leq K \epsilon_n.
\end{align}
Thus $h(\mathcal{R}_n^{\text{Tx}}, \bar{\mathcal{R}}_n^{\text{Tx}}) \leq K \epsilon_n$ 
and $\lim \limits_{n \rightarrow \infty } h(\mathcal{R}_n^{\text{Tx}}, \bar{\mathcal{R}}_n^{\text{Tx}}) = 0$. Hence, $\liminf  \limits_{n \rightarrow \infty} \mathcal{R}_n^{\text{Tx}} = \liminf  \limits_{n \rightarrow \infty} \bar{\mathcal{R}}_n^{\text{Tx}}$. This concludes that 
\begin{align}
\mathcal{C}^{\text{Tx}} = \lim \limits_{n \rightarrow \infty} \mathcal{R}_n^{\text{Tx}}.
\label{ctxfinal}
\end{align}
\begin{remark} 
We can show that the capacity region derived in 
\eqref{ctxfinal} is closed and convex using Lemma 23 and Corollary 24 of \cite{permuter1} which we state here:
\end{remark}
\begin{lemma}
Let $A_n$, $n = 1, 2 \ldots,$ be a sequence of bounded sets in $\mathbb{R}^d$ that includes the origin, i.e., $(0, \ldots, 0)$. If $n A_n$ is sup-additive, i.e., for all $n \geq 1$ and all $N > n$
\begin{align}
N A_N \supseteq n A_n + (N-n) A_{N-n}, \label{supadditivecond}
\end{align}
where sum of two sets $A$ and $B$ is defined as $A + B = \{ \mathbf{a} + \mathbf{b} : \mathbf{a} \in A, \mathbf{b} \in B \}$, and multiplication of a set $A$ with a scalar $c$ is defined as $cA = \{ c \mathbf{a} : \mathbf{a} \in A \}$,
then
\begin{align*}
\lim \limits_{n \rightarrow \infty} A_n = \text{cl} \left(  \bigcup \limits_{n \geq 1} A_n \right).
\end{align*}
Furthermore, for a sup-additive sequence, the limit is convex.
\end{lemma}
We now show that the sequence of sets $\mathcal{R}_n^{\text{Tx}}$ satisfies the sup-additive condition \eqref{supadditivecond}. For any $n \geq 1$ and any $N > n$, we take arbitrary $\mathbf{\bar{R}} \in \mathcal{R}_n^{\text{Tx}}$ and $\mathbf{\tilde{R}} \in \mathcal{R}_{N-n}^{\text{Tx}}$. Then there exists $ \bar{P}_{{U}_{\mk}^{n}}=\prod_i \bar{P}_{{U}_{i}^{n}}$ s.t. $\bar{P}_{{U}_{i}^{n}}\in \mathcal{P}_{i}^{n}$ and $ \tilde{P}_{{U}_{\mk}^{N-n}}=\prod_i \tilde{P}_{{U}_{i}^{N-n}}$ s.t. $\tilde{P}_{{U}_{i}^{N-n}}\in \mathcal{P}_{i}^{N-n} \,\,\text{for}\,\, i = 1, \ldots, K$ such that
\begin{align}
& \sum \limits_{i \in \mi} \bar{R}_i \leq \dfrac{1}{n} I(U_{\mi}^{n} ; Y^{n} | U_{\mi^c}^{n}), \label{mi1} \\
& \sum \limits_{i \in \mi} \tilde{R}_i \leq \dfrac{1}{N-n} I(U_{\mi}^{N-n} ; Y^{N-n} | U_{\mi^c}^{N-n}), \label{mi2}
\end{align}
for every $\mi \subseteq \mk $, where the mutual information in \eqref{mi1} is defined on $\bar{P}_{{U}_{\mk}^{n}}$ and the mutual information in \eqref{mi2} is defined on $\tilde{P}_{{U}_{\mk}^{N-n}}$. Therefore, the point $\mathbf{R} = \frac{n}{N}\mathbf{\bar{R}} + \frac{N-n}{N} \mathbf{\tilde{R}}$ satisfies
\begin{align*}
\sum \limits_{i \in \mi} {R}_i \leq \dfrac{1}{N} \left( 
I(U_{\mi}^{n} ; Y^{n} | U_{\mi^c}^{n}) + I(U_{\mi}^{N-n} ; Y^{N-n} | U_{\mi^c}^{N-n})
\right) 
\end{align*}
for every $\mi \subseteq \mk $.
We note that since we start with empty battery in the beginning, we can define $N$-block input distribution $P_{{U}_{\mk}^{N}}$ by concatenating the input distributions $\bar{P}_{{U}_{\mk}^{n}}$ and $\tilde{P}_{{U}_{\mk}^{N-n}}$, i.e., $P_{{U}_{\mk}^{N}} = \bar{P}_{{U}_{\mk}^{n}} \times \tilde{P}_{{U}_{\mk}^{N-n}}$ without violating the energy constraints. For this choice of input distribution, we have the following for any $\mi \subseteq \mk $:
\begin{align}
I(U_{\mi}^{N} ; Y^{N} | U_{\mi^c}^{N}) & \geq I(U_{\mi}^{n} ; Y^{n} | U_{\mi^c}^{N})
+ I(U_{\mi(n+1)}^{N} ; Y_{n+1}^{N} | U_{\mi}^{n}, U_{\mi^c}^{N})
\nn \\ & = I(U_{\mi}^{n} ; Y^{n} | U_{\mi^c}^{n}) + I(U_{\mi(n+1)}^{N} ; Y_{n+1}^{N},  U_{\mi}^{n}, U_{\mi^c}^{n} | U_{\mi^c(n+1)}^{N}) - I(U_{\mi(n+1)}^{N} ;  U_{\mi}^{n},  U_{\mi^c}^{n} | U_{\mi^c(n+1)}^{N}) \nn \\ & \stackrel{(i)}{=}
I(U_{\mi}^{n} ; Y^{n} | U_{\mi^c}^{n}) + I(U_{\mi(n+1)}^{N} ; Y_{n+1}^{N},  U_{\mi}^{n}, U_{\mi^c}^{n} | U_{\mi^c(n+1)}^{N}) \nn \\ & \geq 
I(U_{\mi}^{n} ; Y^{n} | U_{\mi^c}^{n}) + I(U_{\mi(n+1)}^{N} ; Y_{n+1}^{N} | U_{\mi^c(n+1)}^{N}), \label{supa}
\end{align}
where we define $U_{\mi(n+1)}^{N} = \{ U_{i(n+1)}^{N} : i \in \mi \}$ and (i) is due to the independence of $U_{\mi(n+1)}^{N}$ and $(U_{\mi}^{n}, U_{\mi^c}^{N})$. Note that the first mutual information in \eqref{supa} is defined on $\bar{P}_{{U}_{\mk}^{n}}$ and the second mutual information in \eqref{supa} is defined on $\tilde{P}_{{U}_{\mk}^{N-n}}$, therefore this yields that 
\begin{align*}
\sum \limits_{i \in \mi} {R}_i \leq \dfrac{1}{N} I(U_{\mi}^{N} ; Y^{N} | U_{\mi^c}^{N}),
\end{align*}
for every $\mi \subseteq \mk $. Hence, $\mathbf{R} \in \mathcal{R}_N^{\text{Tx}}$ which concludes that the sequence of sets $\mathcal{R}_n^{\text{Tx}}$ satisfies the sup-additive condition \eqref{supadditivecond}.

\subsubsection{Energy Arrival Information at the Transmitters and the Receiver}
We finish the proof of Theorem \ref{theoremcap} by deriving the capacity region $\mathcal{C}^{\text{TxRx}}$. When the energy arrival information is available causally at the receiver as well, we can repeat the steps in the derivation of capacity region $\mathcal{C}^{\text{Tx}}$ in exactly the same manner. Since the receiver now observes 
$E_{\mk}^{n} = \{ E_{1}^{n}, \ldots, E_{K}^{n} \}$ as well, we simply add it alongside $Y^n$. All the arguments still hold, and we note that
for any distributions $P_{{U}_{i}^{n}} \in \mathcal{P}_i^n$ for $i = 1, \ldots, K$ and any $ \mi \subseteq \{ 1, \ldots, K \} $
\begin{align}
\dfrac{1}{n} I(U_{\mi}^{n};Y^{n}, E_{\mk}^{n} | U_{\mi^c}^{n}) &
 =   \dfrac{1}{n} I(U_{\mi}^{n}; E_{\mk}^{n} | U_{\mi^c}^{n}) + \dfrac{1}{n} I(U_{\mi}^{n};Y^{n}| U_{\mi^c}^{n}, E_{\mk}^{n})
  \nn\\ & \stackrel{(i)}{=} \dfrac{1}{n} I(U_{\mi}^{n};Y^{n}| U_{\mi^c}^{n}, E_{\mk}^{n})
   \nn\\ & =   
   \dfrac{1}{n} 
[ h(Y^{n}|U_{\mi^c}^{n}, E_{\mk}^{n}) 
- h(Y^{n}|U_{\mk}^{n}, E_{\mk}^{n}) ] 
\nn \\ &
\stackrel{(ii)}{=} \dfrac{1}{n} [  
h(Y^{n} | U_{\mi^c}^{n}, E_{\mk}^{n}, X_{\mi^c}^{n})
- h(Y^{n}|U_{\mk}^{n}, E_{\mk}^{n}, X_{\mk}^{n})] 
\nn \\ &
\stackrel{(iii)}{=} \dfrac{1}{n} [
h(Y^{n}| X_{\mi^c}^{n}, E_{\mk}^{n}) 
- h(Y^{n}| X_{\mk}^{n}, E_{\mk}^{n}) ] 
\nn \\ &
=  \dfrac{1}{n} I(X_{\mi}^{n};Y^{n}| X_{\mi^c}^{n}, E_{\mk}^{n})
\label{sum_rsi}
\end{align}
where $(i)$ is because $(U_{1}^{n}, \ldots, U_{K}^{n})$ is independent of $(E_{1}^{n}, \ldots, E_{K}^{n})$, (ii) is due to $X_{i}^{n} = U_{i}^{n}(E_{i}^{n})$ for $i = 1, \ldots, K$, (iii) is because of the Markov chain $(U_{\mi}^{n}) - (X_{\mi}^{n}, E_{\mi}^{n}) - Y^n$ for any $ \mi \subseteq \{ 1, \ldots, K \} $. We note that the final term is based on the distributions $P_{X_{i}^{n} || E_{i}^{n}} \in \mathcal{Q}_{i}^{n}$ on $X_{i}^{n}$ induced by the distributions $P_{{U}_{i}^{n}}$.

Since any distribution $P_{{U}_{i}^{n}} \in \mathcal{P}_i^n$ induces
a distribution $P_{X_{i}^{n} || E_{i}^{n}} \in \mathcal{Q}_{i}^{n}$ on $X_i^n$, and any $X_i^n \sim P_{X_{i}^{n} || E_{i}^{n}} \in \mathcal{Q}_{i}^{n}$ can be represented as random function
of $E_i^n$ according to some distribution $P_{{U}_{i}^{n}} \in \mathcal{P}_i^n$, defining the region
\begin{align*}
 {\mathcal{R}}_n^{\text{TxRx}} =  \bigcup 
\left\{ \mathbf{R} :
\begin{aligned}
& \sum \limits_{i \in \mi} R_i \leq \dfrac{1}{n} I(X_{\mi}^{n} ; Y^{n}|X_{\mi^c}^{n},E_{\mk}^{n}) \\
& \text{for every} \hspace{5pt} \mi \subseteq \mk
\end{aligned}
\right \},
\end{align*}
where the union is over all $P_{X_{\mk}^{n} | E_{\mk}^{n}}$ s.t. $P_{X_{\mk}^{n} | E_{\mk}^{n}}=\prod_i P_{X_{i}^{n} || E_{i}^{n}}$ and $P_{X_{i}^{n} || E_{i}^{n}} \in \mathcal{Q}_{i}^{n} \,\,\text{for}\,\,  i = 1, \ldots, K$, we conclude that the capacity region is 
\begin{align}
\mathcal{C}^{\text{TxRx}} = \lim \limits_{n \rightarrow \infty} {\mathcal{R}}_n^{\text{TxRx}}.
\label{rcapacity}
\end{align}
\begin{remark} 
We can show that the capacity region derived in 
\eqref{rcapacity} is closed and convex following the similar lines as in the previous case.
\end{remark}

\section{Capacity Outer Bound}
\label{sec:outer_bounds}

In this section, we prove Proposition \ref{propObound} and derive the outer bound region for $\mathcal{C}^{\text{Tx}}$ and $\mathcal{C}^{\text{TxRx}}$. We note that since we can always ignore the energy arrival information at the receiver, we have $\mathcal{C}^{\text{Tx}} \subseteq \mathcal{C}^{\text{TxRx}}$. Therefore the outer bound region for $\mathcal{C}^{\text{TxRx}}$  will also hold for $\mathcal{C}^{\text{Tx}}$. 

We fix $n$ and take a rate tuple $(R_1, \ldots, R_K) \in {\mathcal{R}}_n^{\text{TxRx}}$. There exists distributions $ P_{X_{1}^{n} || E_{1}^{n}} \in \mathcal{Q}_{1}^{n}, \ldots, P_{X_{K}^{n} || E_{K}^{n}} \in \mathcal{Q}_{K}^{n}$ such that
\begin{align}
\sum \limits_{i \in \mi} R_i \leq \dfrac{1}{n} I(X_{\mi}^{n} ; Y^{n}|X_{\mi^c}^{n}, E_{\mk}^{n})
\end{align}
for all $\mi \subseteq \{ 1, \ldots, K \}$.
We have
\begin{align}
I(X_{\mi}^{n} ; Y^{n}|X_{\mi^c}^{n}, E_{\mk}^{n})  = 
\sum \limits_{e_{1}^{n}, \ldots, e_{K}^{n}} P_{E_{1}^{n}, \ldots, E_{K}^{n}} ({e}_{1}^{n}, \ldots, {e}_{K}^{n}) I_{{e}_{1}^{n}, \ldots, {e}_{K}^{n}}(X_{\mi}^{n} ; Y^{n}|X_{\mi^c}^{n}),
\label{notation}
\end{align}
where we introduce the notation 
$I_{{e}_{1}^{n}, \ldots, {e}_{K}^{n}}(X_{\mi}^{n} ; Y^{n}|X_{\mi^c}^{n}) \triangleq I(X_{\mi}^{n} ; Y^{n} | X_{\mi^c}^{n}, E_{1}^{n} = {e}_{1}^{n}, \ldots, E_{K}^{n} = {e}_{K}^{n})$. We  note that
\begin{align*}
I_{{e}_{1}^{n}, \ldots, {e}_{K}^{n}}(X_{\mi}^{n} ; Y^{n}|X_{\mi^c}^{n})   & =
  h(Y^{n} | X_{\mi^c}^{n}, E_{1}^{n} = {e}_{1}^{n}, \ldots, E_{K}^{n} = {e}_{K}^{n}) 
- h(Y^{n} | X_{1}^{n}, \ldots, X_{K}^{n},  E_{1}^{n} = {e}_{1}^{n}, \ldots, E_{K}^{n} = {e}_{K}^{n}),
 \\ & \stackrel{(i)}{\leq}
 \sum \limits_{t = 1}^{n} h(Y_{t} | X_{\mi^c t}, E_{1}^{t} = {e}_{1}^{t}, \ldots, E_{K}^{t} = {e}_{K}^{t})  - \sum \limits_{t = 1}^{n}  h(Y_t | X_{1t}, \ldots,  X_{Kt}, E_{1}^{t} = {e}_{1}^{t}, \ldots, E_{K}^{t} = {e}_{K}^{t}),
  \\
 & = 
 \sum \limits_{t = 1}^{n} I_{{e}_{1}^{t},
 \ldots, {e}_{K}^{t}}(X_{\mi t};Y_{t} | X_{\mi^c t}),
\end{align*}
where we define $X_{\mi t} = \{ X_{it} : i \in \mi \}$ and (i) is because conditioning reduces entropy and because fixing $E_1^n = e_1^n, \ldots, E_K^n = e_K^n$ renders the channel  memoryless and because of the structure of $P_{X_{i}^{n} || E_{i}^{n}}$ we only need conditioning on $E_i^t$. We further have
\begin{align}
\dfrac{1}{n} I(X_{\mi}^{n} ; Y^{n}|X_{\mi^c}^{n},E_{1}^{n}, \ldots, E_{K}^{n})   &  \leq \dfrac{1}{n}  \sum \limits_{e_{1}^{n}, \ldots, e_{K}^{n}} P_{E_{1}^{n}, \ldots, E_{K}^{n}} ({e}_{1}^{n}, \ldots, {e}_{K}^{n}) 
\sum \limits_{t = 1}^{n} I_{{e}_{1}^{t},
 \ldots, {e}_{K}^{t}}(X_{\mi t};Y_{t} | X_{\mi^c t})
\nn \\ & \stackrel{(i)}{\leq}
\dfrac{1}{n} \sum \limits_{e_{1}^{n}, \ldots, e_{K}^{n}} P_{E_{1}^{n}, \ldots, E_{K}^{n}} ({e}_{1}^{n}, \ldots, {e}_{K}^{n})   \sum \limits_{t = 1}^{n}
\dfrac{1}{2} \log \left(   1 + \sum \limits_{i \in \mi}  \mathbb{E}\left[   ({X}_{it})^2  | {E}_{i}^{t} = {e}_{i}^{t} \right] \right), \nn \\ & =
\dfrac{1}{n} \mathbb{E}\left[ \sum \limits_{t = 1}^{n}
\dfrac{1}{2} \log \left(   1 + \sum \limits_{i \in \mi}  \mathbb{E}\left[   ({X}_{it})^2  | {E}_{i}^{t}  \right] \right)
\right], \nn \\  & \stackrel{(ii)}{\leq}
\dfrac{1}{2} \log \left(   1 + \dfrac{1}{n} \sum \limits_{t = 1}^{n} \sum \limits_{i \in \mi} \mathbb{E}\left[ \mathbb{E}\left[   ({X}_{it})^2  | {E}_{i}^{t}  \right] \right]  \right),
\nn \\ & =
\dfrac{1}{2} \log \left(   1 + \dfrac{1}{n} \sum \limits_{t = 1}^{n} \sum \limits_{i \in \mi} \mathbb{E}\left[  ({X}_{it})^2 \right] \right), \nn \\ & \stackrel{(iii)}{\leq}
\dfrac{1}{2} \log \left(   1 + \dfrac{1}{n} \sum \limits_{i \in \mi} \mathbb{E}\left[ \sum \limits_{t = 1}^{n} E_{it} \right] \right), \nn \\ & =
\dfrac{1}{2} \log \left(   1  +  \sum \limits_{i \in \mi} \mathbb{E}\left[ E_i \right] 
\right),
\end{align}
where (i) is because $I(X_{\mi};X_{1}+ \ldots, + X_{K} +N | X_{\mi^c}) = 
I(X_{\mi}; X_{\mi} + N)$ is maximized when $X_{i} \sim \mathcal{N}(0, \mathbb{E} X_{i}^{2})$ for $i \in \mi$ since $X_{1t}, \ldots, X_{Kt}$ are independent given ${E}_{1}^{t}, \ldots,  {E}_{K}^{t}$ and (ii) is from Jensen's inequality and (iii) is due to the energy
feasibility constraints \eqref{encos} and \eqref{encos2}. 

We conclude that the rate tuple $(R_1, \ldots, R_K)$ satisfies
\begin{align}
\sum \limits_{i \in \mi} R_i \leq 
\dfrac{1}{2} \log \left(   1  +  \sum \limits_{i \in \mi} \mathbb{E}\left[ E_i \right] 
\right)
\label{d3}
\end{align}
for all $\mi \subseteq \{ 1, \ldots, K \}$.

Note that for fixed $n$, since \eqref{d3} holds for any rate
tuple $(R_1, \ldots, R_K) \in {\mathcal{R}}_n^{\text{TxRx}}$, defining the
region:
\[
\overline{\mathcal{C}}=\left\{
\mathbf{R}:
\begin{aligned}
& \sum \limits_{i \in \mi} R_i \leq \dfrac{1}{2} \log \left( 1 + \sum \limits_{i \in \mi} \mathbb{E}\left[ E_i \right]  \right)  \\
& \text{for every} \hspace{5pt} \mi \subseteq \mk
\end{aligned}
\right\},
\]
it follows that ${\mathcal{R}}_n^{\text{TxRx}} \subseteq \overline{\mathcal{C}} $. Therefore, $\mathcal{C}^{\text{Tx}} \subseteq \mathcal{C}^{\text{TxRx}} = \lim \limits_{n \rightarrow \infty} {\mathcal{R}}_n^{\text{TxRx}} \subseteq   \overline{\mathcal{C}}$.

\section{Throughput Region Bounds}
\label{sec:powercontrolMAC}

In this section, we prove Theorem \ref{theoremthroughput} and derive the inner and outer bound regions for the long-term average throughput region of the energy harvesting MAC. We start with the derivation of the outer region. For any $n$, any set of policies $\mathbf{g}_1, \ldots, \mathbf{g}_K$, and any subset $\mi \subseteq \{1, 2, \ldots, K\}$, we have
\begin{align}
\dfrac{1}{n} \sum 
\limits_{t = 1}^{n} \mathbb{E} \left[
\dfrac{1}{2} \log\left( 1 + \sum_{i \in \mi} g_{it}(E_i^t)\right)  
\right]  & \stackrel{(i)}{\leq} 
\dfrac{1}{2} \log \left( 
1 + \dfrac{1}{n} \mathbb{E} \left[ 
\sum \limits_{t = 1}^{n} \sum_{i\in \mi} g_{it}(E_i^t)
\right] \right) \nn
\\ & \stackrel{(ii)}{\leq}
\dfrac{1}{2} \log \left( 
1 + \dfrac{1}{n} \mathbb{E} \left[ \sum \limits_{t = 1}^{n}  \sum_{i\in \mi}   
E_{it}  
\right] \right) \nn
\\ & = 
\dfrac{1}{2} \log \left( 
1 +   \sum_{i\in \mi}
\mathbb{E} [E_{i}]
\right) \nn
\end{align}
where (i) is from Jensen's inequality and (ii) is due to the energy 
feasibility constraints \eqref{encos} and \eqref{encos2}. Since the inequalities hold for any set of policies $\mathbf{g}_1, \ldots, \mathbf{g}_K$, and any subset $\mi \subseteq \{1, 2, \ldots, K\}$, it follows that $\mathcal{T}_{n} \subseteq \overline{\mathcal{C}} $. Therefore, $\mathcal{T} = \liminf \limits_{n \rightarrow \infty} \mathcal{T}_{n} \subseteq \overline{\mathcal{C}}$.

We continue with the derivation of the inner region. We let the users independently apply the single-user strategy of \cite{dorarxiv},
i.e., the fixed fraction policy. Therefore, denoting $q_i \triangleq \mu_i/\bar{B}_i$ for $i = 1, \ldots, K$, at time $t$ the transmitter $i$ allocates energy
\begin{align}
g_{it} = q_i b_{it}.
\end{align}
For this choice of policies $\mathbf{g}_1, \ldots, \mathbf{g}_K$, 
by definition, we have $\mathcal{T}(g_{1}^{n}, g_{2}^{n}, \ldots, g_{K}^{n}) \subseteq \mathcal{T}_n$, hence
\begin{align*}
\liminf \limits_{n \rightarrow \infty} \mathcal{T}(g_{1}^{n}, g_{2}^{n}, \ldots, g_{K}^{n}) \subseteq \liminf \limits_{n \rightarrow \infty} \mathcal{T}_{n} = \mathcal{T}.
\end{align*}

The fixed fraction policy provides the following result  in \cite[Theorem 2]{dorarxiv}:
\begin{proposition} \label{fixedfractionpolicy}
The fixed fraction policy $\mathbf{g}$ constructed in \cite[Theorem 2]{dorarxiv} satisfies 
\begin{align} 
 \liminf \limits_{n\to\infty}\frac{1}{n}\mathbb{E}\left[\sum_{t=1}^{n}	\frac{1}{2}\log(1+\alpha g_{t}(E^t))\right] 
 \geq
\frac{1}{2}\log(1+\alpha\mathbb{E}[E])- 0.72.
\label{eq:ffp}
\end{align}
\end{proposition}
For any $\mi \subseteq \mk$, we define
\begin{align*}
\lambda_i = \frac{\mathbb{E}[E_i]}{\sum \limits_{j \in \mi} \mathbb{E}[E_j] },
\end{align*}
for $i \in \mi$. We note that $\sum_{i \in \mi} \lambda_i = 1$.
By concavity, we have for every~$t$, 
\begin{align*}
& \frac{1}{2}\log\left( 1 + \sum \limits_{i \in \mi} g_{it}(E_i^t) \right) \geq \sum \limits_{i \in \mi} \lambda_i \frac{1}{2}\log(1+\lambda_i^{-1}g_{it}(E_i^t)).
\end{align*}
Therefore,
\begin{align}
\frac{1}{n} \mathbb{E}\left[ \sum_{t=1}^n \frac{1}{2} \log\left(1+\sum_{k\in \mi} g_{kt}(E_k^t)\right ) \right]  \geq \sum \limits_{i \in \mi} \lambda_i \frac{1}{n} \mathbb{E}\left[ \sum_{t=1}^n \frac{1}{2} \log\left(1+ \lambda_i^{-1} g_{it}(E_i^t)\right ) \right].
\label{eq:ffpset}
\end{align}
Considering \eqref{eq:ffpset} for $n\to\infty$ with Proposition \ref{fixedfractionpolicy} yields
\begin{align}
\liminf \limits_{n\to\infty} \frac{1}{n} \mathbb{E}\left[ \sum_{t=1}^n \frac{1}{2} \log\left(1+\sum_{i\in \mi} g_{it}(E_i^t)\right ) \right] 
& \geq \sum \limits_{i \in \mi}  \lambda_i \frac{1}{2} \log \left(  1 + \lambda_i^{-1} \mathbb{E}\left[ E_i \right] \right)   - 0.72, \nn \\ & =
\dfrac{1}{2} \log\left( 1 + \sum \limits_{i \in \mi}  \mathbb{E}\left[ E_i \right]  \right) - 0.72, \label{tliminf1}
\end{align}
where the equality is because $\lambda_i = \mathbb{E}[E_i]/\sum \limits_{j \in \mi} \mathbb{E}[E_j]$. By definition,
$\forall \epsilon > 0$, $\exists n_0 \in \mathbb{N}$ such that $\forall n \geq n_0$
\begin{align}
\liminf \limits_{n\to\infty} \frac{1}{n} \mathbb{E}\left[ \sum_{t=1}^n \frac{1}{2} \log\left(1+\sum_{i\in \mi} g_{it}(E_i^t)\right ) \right]  < \frac{1}{n} \mathbb{E}\left[ \sum_{t=1}^n \frac{1}{2} \log\left(1+\sum_{i\in \mi} g_{it}(E_i^t)\right ) \right] + \epsilon. \label{tliminf2}
\end{align}
Then from \eqref{tliminf1} and \eqref{tliminf2}, defining the region 
\[
\underline{\mathcal{C}}_\epsilon=\left\{
\mathbf{R}:
\begin{aligned}
& \sum \limits_{i \in \mi} R_i \leq \dfrac{1}{2} \log \left( 1 + \sum \limits_{i \in \mi}  \mathbb{E}\left[ E_i \right]  \right) - 0.72 - \epsilon  \\
& \text{for every} \hspace{5pt} \mi \subseteq \mk
\end{aligned}
\right\},
\]
we have $\underline{\mathcal{C}}_\epsilon \subseteq \liminf \limits_{n \rightarrow \infty} \mathcal{T}(g_{1}^{n}, g_{2}^{n}, \ldots, g_{K}^{n}) \subseteq \mathcal{T}$. Since $\epsilon$ is arbitrary, we obtain
$
\underset{\epsilon > 0}{\bigcup} \underline{\mathcal{C}}_\epsilon \subseteq \mathcal{T}.
$
With a similar approach, we can show that the sequence of sets $\mathcal{T}_n$ satisfies the sup-additive condition \eqref{supadditivecond}, therefore, $\mathcal{T}$ is closed. Hence, we conclude that $\underline{\mathcal{C}}(0.72)  \subseteq \mathcal{T} $.

\section{Capacity Inner Bounds}
\label{sec:lower_bounds}

\subsection{Proof of Theorem \ref{theorem2}}

We start with deriving the inner bound to $\mathcal{C}^{\text{TxRx}}$.
We take some online policies $\mathbf{g}_1\in\mathcal{G}_1, \ldots, \mathbf{g}_K\in\mathcal{G}_K$ and we fix $n$. At time $t = 1, \ldots, n$, we transmit symbols chosen from the uniform distributions on the intervals $\left[ -\sqrt{g_{it}(E_{i}^{t})}, \sqrt{g_{it}(E_{i}^{t})}\right] $ for $i = 1, \ldots, K$. We construct input distributions of the form
$
P_{X_{i}^{n} || E_{i}^{n}}(x_{i}^{n} || e_{i}^{n}) = \prod \limits_{t = 1}^{n}
P_{X_{it} |  E_{i}^{t}}(x_{it} |   e_{i}^{t}),
$
where ${X_{it} |  E_{i}^{t}} \sim \text{Unif} \left( -\sqrt{g_{it}(E_{i}^{t})}, \sqrt{g_{it}(E_{i}^{t})} \right) $. Note that $X_{it}$'s are independent given $E_{i}^{n}$, and since 
$X_{it}^2 \leq g_{it}(E_{i}^{t})$, the energy constraints are satisfied completely by $g_{it}$. We further note that
$X_{1t}, \ldots, X_{Kt}$ are independent  given $(E_{1}^{t}, \ldots, E_{K}^{t})$.  In this case, for every $\mathcal{I} \subseteq \mathcal{K} $ we have
\begin{align}
I(X_{\mathcal{I}}^{n};Y^{n}| X_{\mathcal{I}^c}^{n}, E_{\mathcal{K}}^{n}) & =  \sum \limits_{e_{1}^{n}, \ldots, e_{K}^{n}} P_{E_{1}^{n}, \ldots, E_{K}^{n}} ({e}_{1}^{n}, \ldots, {e}_{K}^{n})  I_{e_{1}^{n}, \ldots, e_{K}^{n}}(X_{\mathcal{I}}^{n};Y^{n}| X_{\mathcal{I}^c}^{n}),  
\nn \\ & =  \sum \limits_{e_{1}^{n}, \ldots, e_{K}^{n}} P_{E_{1}^{n}, \ldots, E_{K}^{n}} ({e}_{1}^{n}, \ldots, {e}_{K}^{n}) \sum \limits_{t = 1}^{n} I_{e_{1}^{n}, \ldots, e_{K}^{n}}(X_{\mathcal{I} t} ; Y_t | X_{\mathcal{I}^c t} ).
\label{lowerb1}
\end{align}
The entropy power inequality yields
\begin{align}
I_{e_{1}^{n}, \ldots, e_{K}^{n}}(X_{\mathcal{I} t} ; X_{1t} + \ldots + X_{Kt} + N_t | X_{\mathcal{I}^c t} )
 & = 
h\left( \sum \limits_{i \in \mathcal{I}} X_{it} + N_t \middle | E_1^t = e_1^t, \ldots,  E_K^t = e_K^t \right)  - h(N_t)
\nn \\ & \geq \dfrac{1}{2} \log \left(
\sum \limits_{i \in \mathcal{I}}  2^{2h(X_{it})} + 2^{2h(N_{t})} \right)    - h(N_t) \nn \\ & = \dfrac{1}{2} \log\left( \sum \limits_{i \in \mathcal{I}} 4 g_{it}(e_{i}^{t}) + 2 \pi e \right) - \dfrac{1}{2} \log\left( 2 \pi e \right) 
\nn \\ & = 
\dfrac{1}{2} \log\left( 1 +  \dfrac{ \sum \limits_{i \in \mathcal{I}} 2 g_{it}(e_{i}^{t}) }{\pi e }\right)
\nn \\ & \geq  \dfrac{1}{2} \log\left( 1 +   \sum \limits_{i \in \mathcal{I}} g_{it}(e_{i}^{t}) \right) - \dfrac{1}{2} \log\left( \dfrac{\pi e}{2}\right),
\label{lowerb2}
\end{align}
where the last inequality is due to the inequality $\log (1 + \alpha x) \geq \log(1+x) + \log \alpha$ for $0 < \alpha \leq 1$. Since $ \frac{1}{2} \log\left( \frac{\pi e}{2}\right) \approx 1.05$, from \eqref{lowerb1} and \eqref{lowerb2} we have
\begin{align*}
\dfrac{1}{n} I(X_{\mathcal{I}}^{n};Y^{n}| X_{\mathcal{I}^c}^{n}, E_{\mathcal{K}}^{n})  \geq    \dfrac{1}{n} \mathbb{E} \left[ \sum \limits_{t = 1}^{n} \dfrac{1}{2} \log\left( 1 +  \sum \limits_{i \in \mathcal{I}} g_{it}(E_{i}^{t}) \right)  \right] - 1.05.
\end{align*}

Denote the finite-horizon throughput of
a set of policies $\mathbf{g}_{\mathcal{I}} = \{ \mathbf{g}_i : i \in \mathcal{I} \}$ up to time $n$ by $\mathscr{T}_n(\mathbf{g}_{\mathcal{I}})=\dfrac{1}{n}\mathbb{E}\left[\sum \limits_{t=1}^{n}\dfrac{1}{2}\log\left( 1 + \sum \limits_{i \in T}  g_{it}(E_i^t)\right) \right]$.
Then
\begin{align}
& \frac{1}{n} I(X_{\mathcal{I}}^{n};Y^{n}| X_{\mathcal{I}^c}^{n},  E_{\mathcal{K}}^{n})
\geq\mathscr{T}_n(\mathbf{g}_{\mathcal{I}})-1.05,
\label{eq:sumrate_g_lower_bound}
\end{align}
for every $\mathcal{I} \subseteq \mathcal{K} $.

For every $\mathbf{g}_1\in\mathcal{G}_{1}, \ldots,  \mathbf{g}_K\in\mathcal{G}_{K}$, define the following region:
\[
\underline{\mathcal{R}}_n^{\text{TxRx}}(\mathbf{g}_1, \ldots, \mathbf{g}_K)
=\left\{\mathbf{R}:
\begin{aligned}
& \sum \limits_{i \in \mathcal{I}} R_i  \leq \mathscr{T}_n(\mathbf{g}_{\mathcal{I}})-1.05 \\
& \text{for every} \hspace{5pt} \mathcal{I} \subseteq \mathcal{K}
\end{aligned}
\right\}.
\]
Then from \eqref{eq:sumrate_g_lower_bound}: $\underline{\mathcal{R}}_n^{\text{TxRx}}(\mathbf{g}_1, \ldots,\mathbf{g}_K)\subseteq
\mathcal{R}_n^{\text{TxRx}}$. Following similar lines as in the proof of Theorem \ref{theoremthroughput} in Appendix \ref{sec:powercontrolMAC}, we conclude that $\underline{\mathcal{C}}^{\text{TxRx}}(\mathbf{g}_1, \ldots,\mathbf{g}_K)\subseteq \mathcal{C}^{\text{TxRx}}$.

We continue with deriving the inner bound to $\mathcal{C}^{\text{Tx}}$.
We construct input distributions $P_{{U}_{i}^{n}} \in \mathcal{P}_{i}^{n}$ according to independent strategy letters:
\begin{align}
P_{{U}_{i}^{n}}({u}_{i}^{n}) = \prod \limits_{t=1}^{n} P_{{U}_{it}}({u}_{it}).
\end{align}
We take some online policies $\mathbf{g}_1\in\mathcal{G}_1, \ldots, \mathbf{g}_K\in\mathcal{G}_K$ and we fix $n$.
We construct $P_{{U}_{it}}$ such that $g_{it}(e_i^t)$ alone will determine the distribution of $X_{it}$, i.e. that $g_{it}(E_i^t)$ will be a sufficient statistic to $X_{it}$. This implies that for two different energy arrival realizations, say $e_i^t$ and $\tilde{e}_i^t$, that satisfy $g_{it}(e_i^t) = g_{it}(\tilde{e}_i^t)$, we must have ${U}_{it}(e_i^t) = {U}_{it}(\tilde{e}_i^t)$ with probability 1. We similarly choose the uniform distribution as in the previous case:  ${U}_{it}(e_i^t) \sim \text{Unif} \left( -\sqrt{g_{it}(e_{i}^{t})}, \sqrt{g_{it}(e_{i}^{t})} \right) $. Let us denote this uniform random variable as $P_{X_i}^{[g_{it}(e_{i}^{t})]}$.

We can think of ${U}_{it}$ as a vector of size $|\mathcal{E}_i|^{t}$ with elements $\{ {U}_{it}(e_i^t), \hspace{5pt} e_i^t \in \mathcal{E}_i^{t} \}$. We will specify the joint distribution of this multivariate random variable. For that matter, define the set of all possible outcomes of the power allocation policy at time $t$:
\begin{align}
\mathcal{O}_{it} = \{    g \in \mathbb{R}_{+} \hspace{5pt}| \hspace{5pt} g =  g_{it}(e_{i}^{t}), \hspace{5pt} e_{i}^{t} \in \mathcal{E}_i^{t}  \}.
\end{align}
We define $P_{{U}_{it}}$ according to the above discussion as:
\begin{align}
P_{{U}_{it}}({u}_{it}) = \prod \limits_{g \in \mathcal{O}_{it}} P_{X_i}^{[g]} ({u}_{it}(g))  \prod \limits_{e_{i}^{t} :  g_{it}(e_{i}^{t}) = g } \mathbbm{1}({u}_{it}({e}_{i}^{t}) = {u}_{it}(g)).
\end{align}
Note that since $g_i^{n} \in \mathcal{G}_{i}$, these input distributions are admissible, i.e. $P_{{U}_{i}^{n}} \in \mathcal{P}_{i}^{n}$.

We proceed to lower bound $I(U_{\mathcal{I}}^{n} ; Y^{n} | U_{\mathcal{I}^c}^{n})$ for any $ \mathcal{I} \subseteq \mathcal{K}$ for these distributions:
\begin{align*}
 I(U_{\mathcal{I}}^{n} ; Y^{n} | U_{\mathcal{I}^c}^{n})
 & = I(U_{\mathcal{I}}^{n} ; Y^{n}, G_{\mathcal{K}}^{n}| U_{\mathcal{I}^c}^{n}) - I(U_{\mathcal{I}}^{n} ; G_{\mathcal{K}}^{n} | Y^{n}, U_{\mathcal{I}^c}^{n}) \\ & 
\stackrel{(i)}{=} I(U_{\mathcal{I}}^{n} ; Y^{n} | U_{\mathcal{I}^c}^{n}, G_{\mathcal{K}}^{n} )  - I(U_{\mathcal{I}}^{n} ; G_{\mathcal{K}}^{n} | Y^{n}, U_{\mathcal{I}^c}^{n}) \\ & \geq I(U_{\mathcal{I}}^{n} ; Y^{n} | U_{\mathcal{I}^c}^{n}, G_{\mathcal{K}}^{n})  -
 H(G_{\mathcal{K}}^{n}),
\end{align*}
where we define $G_{\mathcal{K}}^{n} \triangleq (g_1^{n}(E_1^{n}), \ldots, g_K^{n}(E_K^{n}))$ and (i) is due to $G_{\mathcal{K}}^{n}$ is independent of $U_\mi^{n} = \{ U_i^n : i \in \mi \}$ for any $\mathcal{I} \subseteq \mathcal{K}$. Note that
\begin{align*} 
 I(U_{\mi}^{n} ; Y^{n} | U_{\mi^c}^{n}, G_{\mk}   ) & =
h(Y^{n} | U_{\mi^c}^{n}, G_{\mk} )  - 
h(Y^{n} |  U_{\mk}^{n}, G_{\mk})  \nn \\ & \stackrel{(i)}{=} h(Y^{n} | X_{\mi^c}^{n}, G_{\mk} )  - h(Y^{n} |  X_{\mk}^{n}, G_{\mk} )  \nn \\ & \stackrel{(ii)}{\geq}
h(Y^{n} | X_{\mi^c}^{n}, E_{\mk}^{n} )  - h(Y^{n} |  X_{\mk}^{n}, E_{\mk}^{n} )
\nn \\ & = 
I(X_{\mi}^{n} ; Y^{n} | X_{\mi^c}^{n}, E_{\mk}^{n}),
\end{align*}
where (i) is because $X_{i}^{n} = U_i^{n}(g_i^{n}(E_i^n))$ and the Markov Chain $U_{\mi}^{n} - (X_{\mi}^{n}, G_{\mk}) - Y^n$ for any $\mi \subseteq \mk$ and (ii) is due to the Markov Chain $E_{\mk}^{n} - (G_{\mk}^{n}, X_{\mk}^{n}) - Y^n$ and $g_i^{n}(E_i^n)$ is a deterministic function of $E_i^{n}$.

Therefore, we have the following lower bound for every $ \mi \subseteq \mk$:
\begin{align}
& I(U_{\mi}^{n} ; Y^{n} | U_{\mi^c}^{n}) \geq
I(X_{\mi}^{n} ; Y^{n} | X_{\mi^c}^{n}, E_{\mk}^{n}) -  H(G_{\mathcal{K}}^{n}).
\label{eq:reclower}
\end{align}
The rest follows as before, and we can show that for any $n$ and any sequence of policies $\mathbf{g}_1\in\mathcal{G}_{1}, \ldots, \mathbf{g}_K\in\mathcal{G}_{K}$, we have $\underline{\mathcal{R}}_n^{\text{Tx}}(\mathbf{g}_1, \ldots, \mathbf{g}_K)\subseteq
\mathcal{R}_n^{\text{Tx}}$, where 
$ \underline{\mathcal{R}}_n^{\text{Tx}}(\mathbf{g}_1, \ldots, \mathbf{g}_K)
= $ 
\[ \hspace{-0.08in}
\left\{\mathbf{R}:
\begin{aligned}
& \sum \limits_{i \in \mi} R_i  \leq \mathscr{T}_n(\mathbf{g}_\mi)-
\dfrac{1}{n} H( g_1^{n}(E_1^n), \ldots, g_K^{n}(E_K^n) ) - 1.05\\
& \text{for every} \hspace{5pt} \mi \subseteq \mk
\end{aligned}
\right\}
\] 
Following a similar analysis as in the previous section, we get $\underline{\mathcal{C}}^{\text{Tx}}(\mathbf{g}_1, \ldots,\mathbf{g}_K)\subseteq\mathcal{C}^{\text{Tx}}$.

\subsection{Proof of Theorem \ref{theorem3}}
We apply the result of Theorem \ref{theorem2} with a particular pair of policies. The online policy introduced in \cite[Theorem 3]{maincapacityarticle} provides the following result:
\begin{proposition} \label{onlinepolicy}
The online power control policy $\mathbf{g}$ constructed in \cite[Theorem 3]{maincapacityarticle} satisfies 
\begin{equation} 
\mathscr{T}(\alpha\cdot\mathbf{g})\geq\frac{1}{2}\log(1+\alpha\mathbb{E}[E])-1.80,
\label{eq:Tg1_gap}
\end{equation}
for any scalar $\alpha>0$ where
$\mathscr{T}(\alpha\cdot\mathbf{g})=
\liminf \limits_{n\to\infty}
\mathscr{T}_n(\alpha\cdot\mathbf{g})$ and
\[
{\mathscr{T}_n(\alpha\cdot\mathbf{g})}=\frac{1}{n}\mathbb{E}\left[\sum_{t=1}^{n}\frac{1}{2}\log(1+\alpha\cdot g_{t}(E^t))\right].
\]
Furthermore, its entropy rate is bounded as
\begin{align*}
H(\mathbf{g}(\mathbf{E})) \leq 1.
\end{align*}
\end{proposition}
We note that the proof of \eqref{eq:Tg1_gap} is specific to the case $\alpha = 1$ in \cite[Theorem 3]{maincapacityarticle}, however, it can be trivially extended to any $\alpha>0$. The bound on the entropy rate is
showed in \cite[Section V-C]{maincapacityarticle}.

We consider using the online policy introduced in \cite[Theorem 3]{maincapacityarticle} independently 
in all transmitters. A similar treatment as in the proof of Theorem~\ref{theoremthroughput} yields that for any $\mi \subseteq \mk$
\begin{align}
\mathscr{T}(\mathbf{g}_\mi)
 &\geq 
\dfrac{1}{2} \log\left( 1 + \sum \limits_{i \in \mi}  \mathbb{E}\left[ E_i \right]  \right) - 1.80.
\label{double}
\end{align}
We further obtain the following bound from Proposition
\ref{onlinepolicy}:
\begin{align}
H(\mathbf{g}_1(\mathbf{E}_1), \ldots, \mathbf{g}_K(\mathbf{E}_K)) & \leq 
H(\mathbf{g}_1(\mathbf{E}_1))  + \ldots + H(\mathbf{g}_K(\mathbf{E}_K)) \nn \\ &
\leq K.
\label{entrateK}
\end{align}
Substituting \eqref{double} and \eqref{entrateK} in $\underline{\mathcal{C}}^{\text{Tx}}(\mathbf{g}_1, \ldots,\mathbf{g}_K)$ and applying Theorem \ref{theorem2} implies
$\underline{\mathcal{C}}(2.85 + K)\subseteq \underline{\mathcal{C}}^{\text{Tx}}(\mathbf{g}_1, \ldots,\mathbf{g}_K)
\subseteq
\mathcal{C}^{\text{Tx}}$.

\section{Sum-Capacity of the Energy Harvesting Gaussian MAC}
\label{sec:kuserTxRx}

\subsection{Energy Arrival Information at the Transmitters and the Receiver}

We start with the discussion of submodular functions which we 
will use to prove Propositions \ref{propsym}-\ref{KuserTXbounds} later in this section.

Let $f$ be a set function on a set $\mathcal{K}$, i.e., a function defined on the collection of all subsets of $\mathcal{K}$. The function $f$ is called submodular if
\begin{align}
f(\mi) + f(\mj) \geq f(\mi \cap \mj) + f(\mi \cup \mj)
\end{align}
for all subsets $\mi$, $\mj$ of $\mathcal{K}$. Given any set of distributions $ P_{X_{i}^{n} || E_{i}^{n}} \in \mathcal{Q}_{i}^{n}$ for $i = 1, \ldots, K$, define $f^n(\mi) = I(X_{\mi}^{n}, ;Y^{n}|X_{\mi^c}^{n}, E_{\mathcal{K}}^n)$, where we denote $X_{\mi}^{n} = \{ X_{i}^{n} \hspace{5pt} | \hspace{5pt} i \in \mi \}$, $E_{\mathcal{K}}^n = \{ E_1^n, \ldots, E_K^n \}$ and $\mathcal{K} = \{ 1, \ldots, K  \}$. Using the following theorem, we first show that $f^n$ is a submodular function for every $n \geq 1$ :
\begin{theorem}[Theorem 44.1,\cite{Schrijver:book}] \label{thm:sub}
A set function $f$ on $\mathcal{K}$ is submodular if and only if
\begin{align}
f(\mi \cup \{ s \} ) + f(\mi \cup \{ u \} ) \geq f(\mi) + f(\mi \cup \{ s, u \})
\label{subalternative}
\end{align}
for each $\mi \subseteq \mathcal{K}$ and distinct $s, u \in \mathcal{K} \backslash \mi$. 
\end{theorem}

For simplicity, we define $\mi \cup \{ s \} = \mi_s$, $\mi \cup \{ u \} = \mi_u$ and $\mi \cup \{ s, u \} = \mi_{s, u}$. We have
\begin{align}
I(X_{\mi_s}^{n}; Y^{n} | X_{\mi_s^c}^{n}, E_{\mathcal{K}}^n) - I(X_{\mi}^{n}; Y^{n} | X_{\mi^c}^{n}, E_{\mathcal{K}}^n) & = I(X_{s}^{n}; Y^{n} | X_{\mi_s^c}^{n}, E_\mk^n) + I(X_{\mi}^{n}; Y^{n} | X_{s}^{n}  ,X_{\mi_s^c}^{n}, E_{\mathcal{K}}^n) \nn \\ & \hspace{30pt} - I(X_{\mi}^{n}; Y^{n} | X_{\mi^c}^{n}, E_{\mathcal{K}}^n)\nn \\ & \stackrel{(i)}{=} I(X_{s}^{n}; Y^{n} | X_{\mi_s^c}^{n}, E_{\mathcal{K}}^n),
\label{sub1}
\end{align}
where (i) is due to $\{ X_{s}^{n}  ,X_{\mi_s^c}^{n} \} = \{ X_{\mi^c}^{n} \}$. We further have
\begin{align}
I(X_{\mi_u}^{n}; Y^{n} | X_{\mi_u^c}^{n}, E_{\mathcal{K}}^n) - I(X_{\mi_{s,u}}^{n}; Y^{n} | X_{\mi_{s,u}^c}^{n}, E_{\mathcal{K}}^n)  &  =
I(X_{u}^{n}; Y^{n} | X_{\mi_u^c}^{n}, E_{\mathcal{K}}^n) + I(X_{\mi}^{n}; Y^{n} | X_{u}^{n}  ,X_{\mi_u^c}^{n}, E_{\mathcal{K}}^n) 
\nn \\ & \hspace{20pt} - I(X_{u}^{n}; Y^{n} | X_{\mi_{s,u}^c}^{n}, E_{\mathcal{K}}^n) - I(X_{s}^{n}; Y^{n} |X_{u}^{n} ,  X_{\mi_{s,u}^c}^{n}, E_{\mathcal{K}}^n) \nn \\ & \hspace{35pt} - I(X_{\mi}^{n}; Y^{n} | X_{s}^{n}, X_{u}^{n}, X_{\mi_{s,u}^c}^{n}, E_{\mathcal{K}}^n) \nn \\ & = I(X_{u}^{n}; Y^{n} | X_{\mi_u^c}^{n}, E_{\mathcal{K}}^n)
- I(X_{u}^{n}; Y^{n} | X_{\mi_{s,u}^c}^{n}, E_{\mathcal{K}}^n) \nn \\ & \hspace{20pt} - I(X_{s}^{n}; Y^{n} |X_{u}^{n} ,  X_{\mi_{s,u}^c}^{n}, E_{\mathcal{K}}^n).
\label{sub2}
\end{align}
From \eqref{sub1} and \eqref{sub2}, we get
\begin{align}
& I(X_{\mi_s}^{n}; Y^{n} | X_{\mi_s^c}^{n}, E_{\mathcal{K}}^n ) + I(X_{\mi_u}^{n}; Y^{n} | X_{\mi_u^c}^{n}, E_{\mathcal{K}}^n)  - I(X_{\mi}^{n}; Y^{n} | X_{\mi^c}^{n}, E_{\mathcal{K}}^n)  - I(X_{\mi_{s,u}}^{n}; Y^{n} | X_{\mi_{s,u}^c}^{n}, E_{\mathcal{K}}^n) 
\nn \\ & =
I(X_{s}^{n}; Y^{n} | X_{\mi_s^c}^{n},  E_{\mathcal{K}}^n) + I(X_{u}^{n}; Y^{n} | X_{\mi_u^c}^{n},  E_{\mathcal{K}}^n) 
- I(X_{u}^{n}; Y^{n} | X_{\mi_{s,u}^c}^{n},  E_{\mathcal{K}}^n)  - I(X_{s}^{n}; Y^{n} |X_{u}^{n} ,  X_{\mi_{s,u}^c}^{n},  E_{\mathcal{K}}^n) \nn \\ & =
I(X_{u}^{n}; Y^{n} | X_{\mi_u^c}^{n},  E_{\mathcal{K}}^n)
- I(X_{u}^{n}; Y^{n} | X_{\mi_{s,u}^c}^{n},  E_{\mathcal{K}}^n).
\label{submain}
\end{align}
We further have
\begin{align}
I(X_{u}^{n}; Y^{n} | X_{\mi_u^c}^{n}, E_{\mathcal{K}}^n)
- I(X_{u}^{n}; Y^{n} | X_{\mi_{s,u}^c}^{n}, E_{\mathcal{K}}^n) \nn  & =
I(X_{u}^{n}; Y^{n}, X_{s}^{n} | X_{\mi_{s,u}^c}^{n}, E_{\mathcal{K}}^n) -
I(X_{u}^{n}; X_{s}^{n} | X_{\mi_{s,u}^c}^{n}, E_{\mathcal{K}}^n) \nn \\ & \hspace{30pt} - I(X_{u}^{n}; Y^{n} | X_{\mi_{s,u}^c}^{n}, E_{\mathcal{K}}^n) \nn \\ & \stackrel{(i)}{=}
I(X_{u}^{n}; Y^{n}, X_{s}^{n} | X_{\mi_{s,u}^c}^{n}, E_{\mathcal{K}}^n) 
- I(X_{u}^{n}; Y^{n} | X_{\mi_{s,u}^c}^{n},  E_{\mathcal{K}}^n) \nn \\ & \geq
I(X_{u}^{n}; Y^{n} | X_{\mi_{s,u}^c}^{n},  E_{\mathcal{K}}^n) 
- I(X_{u}^{n}; Y^{n} | X_{\mi_{s,u}^c}^{n},  E_{\mathcal{K}}^n) \nn \\ & = 0
\label{subfinish}
\end{align}
where (i) is due to $X_{i}^{n}$'s are independent given $ E_{\mathcal{K}}^n = \{ E_1^n, \ldots, E_K^n \}$. \eqref{submain} and \eqref{subfinish}
together imply that $f^n$ satisfies \eqref{subalternative} and therefore it is submodular.

We further note that $f(\emptyset) = 0$ and $f(\mi) \leq f(\mj)$ if
$\mi \subseteq \mj$, i.e., $f$ is nondecreasing due to
\begin{align*}
I(X_{\mj}^{n}; Y^{n} | X_{\mj^c}^{n}, E_{\mathcal{K}}^n)  & = I(X_{\mj \backslash \mi}^{n}; Y^{n} | X_{\mj^c}^{n}, E_{\mathcal{K}}^n) + I(X_{\mi}^{n}; Y^{n} | X_{\mi^c}^{n}, E_{\mathcal{K}}^n) \\ & \geq
I(X_{\mi}^{n}; Y^{n} | X_{\mi^c}^{n}, E_{\mathcal{K}}^n).
\end{align*}
Therefore for any $n$ and any input distribution $P_{X_{1}^{n} || E_{1}^{n}} \times \ldots \times P_{X_{K}^{n} || E_{K}^{n}}$, the following region is a polymatroid:
\begin{align}
\left\{
\mathbf{R}:
\begin{aligned}
& \sum_{i\in \mi} R_i \leq \dfrac{1}{n} I(X_{\mi}^{n};Y^{n}|X_{\mi^c}^{n},E_{\mk}^{n}) \\
& \text{for every $\mi \subseteq \mk$.}
\end{aligned}
\right\}
\label{polymatroid}
\end{align}
where a polymatroid associated with a nondecreasing, normalized ($f(\emptyset) = 0$) and submodular set function $f$ on $\mathcal{K}$ is defined as:
\begin{align*}
\{  x \in \mathbb{R}^{K} | \hspace{5pt} x \geq \mathbf{0}, \hspace{5pt} x(\mi) \leq f(\mi) \hspace{5pt} \text{for each} \hspace{5pt} \mi \subseteq \mathcal{K} \},
\end{align*}
where $x(\mi) = \sum \limits_{i \in \mi} x_i$. We continue with the proof of Proposition \ref{propsym}.

\subsubsection{Proof of Proposition \ref{propsym}}

We now focus on the sum-capacity of the energy harvesting Gaussian MAC which we denote by $C_{\text{sum}}$. We will show that
\begin{align}
C_{\text{sum}}^{\text{TxRx}} = \lim \limits_{n \rightarrow \infty} \sup \limits_{\substack{ P_{X_{\mk}^{n} | E_{\mk}^{n}}=\prod_i P_{X_{i}^{n} || E_{i}^{n}}\\  P_{X_{i}^{n} || E_{i}^{n}} \in \mathcal{Q}_{i}^{n}  \\ i = 1, \ldots, K}} \dfrac{1}{n} I(X_{\mk}^{n} ;Y^{n}|E_{\mk}^{n}).
\label{csumtxrxp}
\end{align}
We define $C_n = \sup \left\lbrace  \sum \limits_{i = 1}^{K}  R_i : (R_1, \ldots, R_K) \in {\mathcal{R}}_n^{\text{TxRx}} \right\rbrace $. We note that for any input distribution $P_{X_{1}^{n} || E_{1}^{n}} \times \ldots \times P_{X_{K}^{n} || E_{K}^{n}}$ the corresponding region \eqref{polymatroid} is a polymatroid. It is shown in \cite{Edmonds2003} that for any permutation $\pi$ on set $\mathcal{K}$, the vector $\mathbf{R}(\pi) \in \mathbb{R}^{K}$ defined by
\begin{align*}
&{R}_{\pi(1)}(\pi) = f(\pi(1)), \\
&{R}_{\pi(i)}(\pi) = f(\{
\pi(1), \ldots, \pi(i)\}) - f(\{
\pi(1), \ldots, \pi(i-1)\})
\end{align*}
for $i = 2, \ldots, K$ is a vertex for the corresponding polymatroid. We can observe that for any permutation $\pi$ on set $\mathcal{K}$, the corresponding rate tuple $\mathbf{R}(\pi)$ satisfies 
\begin{align*}
\sum \limits_{i = 1}^{K}  R_i(\pi) = \dfrac{1}{n} I(X_{\mk}^{n};Y^{n}|E_{\mk}^{n}).
\end{align*}
Therefore, for any input distribution $P_{X_{1}^{n} || E_{1}^{n}} \times \ldots \times P_{X_{K}^{n} || E_{K}^{n}}$, the sum-capacity of the corresponding region \eqref{polymatroid} is $\dfrac{1}{n} I(X_{\mk}^{n};Y^{n}|E_{\mk}^{n})$. We can take the supremum over $P_{X_{i}^{n} || E_{i}^{n}} \in \mathcal{Q}_{i}^{n}$ for  $i = 1, \ldots, K$ and obtain
\begin{align}
C_n = \sup \limits_{\substack{ P_{X_{\mk}^{n} | E_{\mk}^{n}}=\prod_i P_{X_{i}^{n} || E_{i}^{n}}\\  P_{X_{i}^{n} || E_{i}^{n}} \in \mathcal{Q}_{i}^{n}  \\ i = 1, \ldots, K}}  \dfrac{1}{n} I(X_{\mk}^{n} ;Y^{n}|E_{\mk}^{n}).
\label{Cnsum}
\end{align}
We first observe that since 
\begin{align*}
\mathcal{C}^{\text{TxRx}} = \liminf \limits_{n \rightarrow \infty} \mathcal{R}_n^{\text{TxRx}} = \bigcup \limits_{n \geq 1}  \bigcap \limits_{m \geq n}\mathcal{R}_m^{\text{TxRx}},
\end{align*}
for any rate tuple $(R_1, \ldots, R_K) \in \mathcal{C}^{\text{TxRx}} $, there exists $n \in \mathbb{N}$ such that $\forall m \geq n$, $(R_1, \ldots, R_K) \in {\mathcal{R}}_m^{\text{TxRx}}$. This implies that $\sum \limits_{i = 1}^{K}  R_i \leq C_m$, $\forall m \geq n$ hence $\sum \limits_{i = 1}^{K}  R_i \leq \liminf \limits_{n \rightarrow \infty} C_n$. Since $(R_1, \ldots, R_K)$ is arbitrary, we have
\begin{align*}
C_{\text{sum}}^{\text{TxRx}} \leq \liminf \limits_{n \rightarrow \infty} C_n.
\end{align*}
On the other hand, since 
 \begin{align*}
\mathcal{C}^{\text{TxRx}} = \limsup \limits_{n \rightarrow \infty} \mathcal{R}_n^{\text{TxRx}} = \bigcap \limits_{n \geq 1}  \bigcup \limits_{m \geq n}\mathcal{R}_m^{\text{TxRx}},
\end{align*}
we have
\begin{align*}
C_{\text{sum}}^{\text{TxRx}} & = \sup\left\lbrace  \sum \limits_{i = 1}^{K}  R_i : (R_1, \ldots, R_K) \in
\bigcap \limits_{n \geq 1}  \bigcup \limits_{m \geq n}\mathcal{R}_m^{\text{TxRx}} \right\rbrace  \\ & \stackrel{(i)}{=}
\inf \limits_{n \geq 1} \sup\left\lbrace  \sum \limits_{i = 1}^{K}  R_i : (R_1, \ldots, R_K) \in
\bigcup \limits_{m \geq n}\mathcal{R}_m^{\text{TxRx}} \right\rbrace 
\\ & =
\inf \limits_{n \geq 1}  \sup \limits_{m \geq n} 
\sup\left\lbrace  \sum \limits_{i = 1}^{K}  R_i : (R_1, \ldots, R_K) \in
\mathcal{R}_m^{\text{TxRx}} \right\rbrace  \\ & =
\limsup \limits_{n \rightarrow \infty} C_n.
\end{align*}
where (i) is because $\bigcup \limits_{m \geq n}\mathcal{R}_m^{\text{TxRx}}$ is decreasing in $n$. This concludes the proof for \eqref{csumtxrxp}.

\subsubsection{Proof of Proposition \ref{KuserTXRXbounds}} We derive here the bounds on the sum-capacity $C_{\text{sum}}^{\text{TxRx}}$ in \eqref{eq:symC1}. We note that the outer bound in Proposition \ref{propObound} gives
the following upper bound on $C_{\text{sum}}^{\text{TxRx}}$ when we take 
$\mi = \mk$: 
\begin{align*}
C_{\text{sum}}^{\text{TxRx}} \leq \dfrac{1}{2} \log \left( 1 + K \mathbb{E}\left[ E \right]  \right).
\end{align*}
For the lower bound, we can use the steps in Theorem \ref{theoremthroughput} and Theorem \ref{theorem2}. If we use the input distributions introduced in Theorem \ref{theorem2} with the online policy introduced in \cite[Theorem 2]{dorarxiv}  independently 
in all transmitters, the inequalities \eqref{lowerb1}-\eqref{eq:sumrate_g_lower_bound} and \eqref{tliminf1} yield the following lower bound on $C_{\text{sum}}^{\text{TxRx}}$ when we take 
$\mi = \mk$: 
\begin{align*}
\dfrac{1}{2} \log \left( 1 + K \mathbb{E}\left[ E \right]  \right)- 1.77\leq C_{\text{sum}}^{\text{TxRx}}
\end{align*} 
which completes the proof of \eqref{eq:symC1}.

\subsection{Energy Arrival Information at the Transmitters Only}

We note that by defining $f^n(\mi) = I(U_{\mi}^{n} ;Y^{n}|U_{\mi^c}^{n})$ for any distributions  $ P_{U_{i}^{n}} \in \mathcal{P}_{i}^{n}$ 
for $i = 1, \ldots, K$, we can similarly show that $f^n$ is a submodular function. This allows us to conclude that the following region is a polymatroid:
\begin{align}
\left\{
\mathbf{R}:
\begin{aligned}
& \sum_{i\in \mi} R_i \leq \dfrac{1}{n} I(U_{\mi}^{n};Y^{n}|U_{\mi^c}^{n}) \\
& \text{for every $\mi \subseteq \mk$.}
\end{aligned}
\right\}.
\label{polymatroid2}
\end{align}
We can finally follow the approach in the previous section and conclude that 
\begin{align*}
C_{\text{sum}}^{\text{Tx}} = \lim \limits_{n \rightarrow \infty}  \sup \limits_{\substack{{P_{{U}_{\mk}^{n}}=\prod_i P_{{U}_{i}}^{n}} \\ P_{{U}_{i}^{n}}\in \mathcal{P}_{i}^{n}  \\ i = 1, \ldots, K  }}  \dfrac{1}{n} I(U_{\mk}^{n};Y^{n}).
\end{align*}

We next derive the bounds on the sum-capacity $C_{\text{sum}}^{\text{Tx}}$ in \eqref{eq:last}. From Proposition \ref{propObound}, we get the following upper bound on $C_{\text{sum}}^{\text{Tx}}$:
\begin{align*}
C_{\text{sum}}^{\text{Tx}} \leq \dfrac{1}{2} \log \left( 1 + K \mathbb{E}\left[ E \right]  \right).
\end{align*}
For the lower bound, we use the inequalities \eqref{eq:sumrate_g_lower_bound} and \eqref{eq:reclower} by taking $\mi = \mk$ and if we use the input distributions introduced in Theorem \ref{theorem2} with the online policy introduced in \cite[Theorem 3]{maincapacityarticle}  independently 
in all transmitters, the inequalities \eqref{double} and \eqref{eq:entropy2} yield the following lower bound on $C_{\text{sum}}^{\text{Tx}}$:
\begin{align*}
\dfrac{1}{2} \log\left( 1 +  K \mathbb{E}\left[ E \right]  \right) -  3.85  
\leq C_{\text{sum}}^{\text{Tx}}
\end{align*}
which completes the proof of \eqref{eq:last}.

\end{document}